\documentclass[prb,twocolumn,showpacs,preprintnumbers,
amsmath,amssymb,superscriptaddress]{revtex4}

\usepackage{graphicx}
\usepackage{dcolumn}
\usepackage{bm}
\newcommand{\beq}{\begin{equation}}
\newcommand{\eeq}{\end{equation}}
\newcommand{\beqa}{\begin{eqnarray}}
\newcommand{\eeqa}{\end{eqnarray}}

\newcommand{\w}{\omega}

\newcommand{\half}{\mbox{$\textstyle \frac{1}{2}$}}
\newcommand{\ket}[1]{\left| #1 \right\rangle}
\newcommand{\bra}[1]{\left\langle #1 \right|}

\newcommand{\upket}{\ket{\uparrow}}
\newcommand{\downket}{\ket{\downarrow}}
\newcommand{\raiz}{\mbox{$\textstyle \frac{1}{\sqrt{2}}$}}

\begin{document}

\preprint{V100603}

\title{Optical Schemes for Quantum Computation in Quantum Dot Molecules }
\author{Brendon~W.~Lovett}
\email{brendon.lovett@materials.oxford.ac.uk}
\affiliation{Department of Materials, Oxford University, Oxford OX1 3PH, United Kingdom}
\author{John~H.~Reina}
\altaffiliation[On leave of absence from ]{Centro Internacional de F\'{\i}sica (CIF), A.A. 4948, Bogot\'a, Colombia}
\email{j.reina-estupinan@physics.oxford.ac.uk}
\affiliation{Department of Materials, Oxford University, Oxford OX1 3PH, United Kingdom}
\affiliation{Centre for Quantum Computation, Clarendon Laboratory, Department of Physics, Oxford University, Oxford OX1 3PU, United Kingdom}
\author{Ahsan~Nazir}
\affiliation{Department of Materials, Oxford University, Oxford OX1 3PH, United Kingdom}
\author{G.~Andrew~D.~Briggs}
\affiliation{Department of Materials, Oxford University, Oxford OX1 3PH, United Kingdom}
\date{\today}
\begin{abstract}
We give three methods for
entangling quantum states in quantum dots. We do this by showing how to tailor the resonant energy (F\"orster-Dexter)
transfer mechanisms and the biexciton binding energy in a quantum dot molecule.
We calculate the magnitude of these two electrostatic interactions as a function of dot size, interdot separation, material
composition, confinement potential and applied electric field by using an envelope function approximation in a two-cuboid
dot molecule. In the first implementation, we show that it is desirable to suppress the F\"orster coupling and to create
entanglement by using the biexciton energy alone. We show how to perform universal quantum logic in a second implementation
which uses the biexciton energy together with appropriately tuned laser pulses:
by selecting appropriate materials parameters high fidelity logic can be achieved.
The third implementation proposes generating quantum entanglement by switching the F\"orster
interaction itself. We show that the energy transfer can be fast enough in certain dot structures that switching can occur
on a timescale which is much less than the typical decoherence times.

\end{abstract}
\pacs{03.67.Lx, 03.67-a, 78.67.Hc, 73.20.Mf }

\maketitle

\section{\label{intro}Introduction}

Quantum dots~\cite{harrison00,bimberg00} are quantum heterostructures which are composed of
nanoscale regions of one type of material which is embedded in a second type. In a semiconductor quantum dot (QD),
materials with differing bandgaps are used; this leads to the possibility of electronic confinement within the dot region.
Moreover, the confined electronic states can be accurately
controlled by varying the dot size, shape or composition, and the number of
confined electrons; all of these may be altered by using different growth conditions and hence
specifically tailored ``artificial atoms" or ``superatoms" can be produced.~\cite{harrison00,bimberg00}
Some prominent atom-like properties of QDs include an electronic
shell structure,~\cite{hawrylak01} Rabi oscillations,~\cite{kama01} photon
antibunching,~\cite{michler00a,rege01} controlled quantum light emission,~\cite{michler00b,santo01}
and quantum entanglement.~\cite{jhthesis,reina00}
One of the most intriguing possible applications of quantum dots is that they may be used to build quantum
computers.~\cite{steane98,nielsen00}
A practical realization of a quantum computer would be very significant, since there
exist theoretical quantum algorithms which would make some classically hard computational problems tractable.~\cite{shor95}
Such quantum devices could also accurately simulate any physical system
(and the evolution of its local interactions) by invoking the same amount of energy and Hilbert space requirements
as the system itself.~\cite{feynman82,abrams97,zalka98}

The basic unit of a quantum computer is a two-level quantum system, the so-called qubit.
Of the utmost importance is the identification of a physical system where a coherent qubit evolution
can be performed, thus allowing a precise execution of the elementary quantum gates required for universal quantum
computation.~\cite{steane98,nielsen00}
Many different types of hardware for embodying qubits have been proposed (for a collection of papers detailing some of these
see Ref.~\onlinecite{review2}) and some of them
have already been implemented for performing elementary quantum gate operations.
These include ion
traps,~\cite{cirac95,cirac00,monroe95,molmer99,sackett00} quantum electrodynamics
cavities,~\cite{pelli95,turche95,cirac96,cirac97,rausch01}
nuclear magnetic resonance,~\cite{chuang97,chuang98,cory97,mosca98},
dopants in semiconductors,~\cite{kane98,vrijen00,privman01}
optical lattices and Bose-Einstein condensates,~\cite{bre99,jack99,greiner02}
Josephson junctions,~\cite{averin98,schon01,naka99,wal00} and quantum
dots.~\cite{bare95,loss98,reina00,reina00b,eliana00,lovettprl,derinaldis02, dassarma00, chen00, elaine03, troiani00}
In this article we concentrate on a quantum dot implementation. Previous proposals~\cite{jhthesis,reina00,bare95,loss98,reina00b,pazy01,biolatti02,lovettprl,derinaldis02, troiani00}
include the use of a single electron~\cite{loss98,pazy01, troiani00} or nuclear~\cite{reina00b} spin located on each of an array of interacting dots, the
presence or absence of an electron charge state,~\cite{bare95} or the use of excitonic
states.~\cite{reina00,biolatti02,lovettprl,derinaldis02} We show how an energy selective approach to manipulating the excitonic states of coupled QDs,
together with control over the energy transfer and biexciton binding energy, can be used to perform quantum computation (QC) and to produce controlled exciton quantum entanglement.
In so doing, we investigate the F\"orster-Dexter resonant energy transfer, a mechanism
first studied in the context of the sensitized luminescence of
solids,~\cite{forster59,dexter53} in which an excited sensitizer atom can transfer its excitation to a neighbouring
acceptor atom, via an intermediate virtual photon.
This mechanism is also responsible for photosynthetic energy processes in antenna complexes,
biosystems (BSs) that harvest sunlight.~\cite{hu02}
More recently, interest has focussed on energy transfer in quantum dot nanostructures~\cite{crooker02}
and within molecular systems (MSs).~\cite{hettich02} In this article we show how to exploit such energy transfer
mechanisms with a view to processing quantum information. This article is a more detailed account of the work which
appears in Ref.~\onlinecite{lovettprl}.

\section{building quantum logic gates}
\label{heisenberg}
We consider the Hamiltonian of two interacting quantum dots. We assume that
the dots are sufficiently far apart that tunnelling processes between them may be neglected
but that there is a strong exciton-exciton coupling. Our two-level system is represented in each dot by
a single low lying exciton state $\ket{1}$ and the ground state $\ket{0}$.
Then the interaction Hamiltonian can be written
in the computational basis ($\{|00\rangle, |01\rangle,|10\rangle,|11\rangle\}$, with the first digit referring
to dot I and the second to dot II) as follows:
\beqa
\widehat{H}=
\left(
\begin{array}{cccc}
\w_0 & 0 & 0 &0 \\
0 & \w_0+\w_2 & V_{\rm F} & 0 \\
0 & V_{\rm F} &  \w_0+\w_1 &0 \\
0 & 0 & 0 &  \w_0+\w_1+\w_2+V_{\rm XX}
\end{array}\right).
\label{eq:Hdot}
\eeqa
The diagonal interaction $V_{\rm XX}$ is the direct Coulomb binding energy
between two excitons, one located on each dot, and $V_{\rm F}$ denotes the Coulomb exchange (F\"orster)
interaction which is off-diagonal and therefore induces the
transfer of an exciton from one QD to the other. These are the only Coulomb interaction terms which act between the qubits
and will be calculated and discussed in detail in Section~\ref{inter}.
$\w_0$ denotes the ground state energy, $\w_1$ ($\w_2$) refers to the energy required to create an exciton on dot I (II)
in the absence of interactions, and includes intra-dot coupling contributions (direct Coulomb binding energy and spin
splitting) which we shall discuss in
Section~\ref{intra}. We also define $\Delta_0 \equiv \w_1-\w_2$ to be the difference between the exciton
creation energy for dot I and that for dot II in the absence of interactions between the dots.
Thus, if $H_0=H_{\rm I}+H_{\rm II}$ denotes the free particle
Hamiltonian, then $H_0(\ket{\gamma_1}\ket{\gamma_2})=(\gamma_1\w_1+\gamma_2\w_2)\ket{\gamma_1}\ket{\gamma_2}$,
$\gamma_1,\gamma_2=0,1$ ($\hbar=1$ throughout this article), then $H=H_0+V_{\rm 1,2}$ (where $V_{\rm 1,2}$ accounts for the
qubit-qubit interactions, $V_{\rm XX}$ and $V_{\rm F}$) is the system's overall Hamiltonian.
In the case of an $n$-qubit register with nearest neighbour interactions,
the Hamiltonian takes the form $H=H_0^{(n)}+\sum_{i=1}^{n-1} V_{i,i+1}$, where
$H_0^{(n)}\equiv \sum_{i=1}^n H_i$ is the free particle Hamiltonian, and $V_{i,i+1}$ are the interaction terms.
A related Hamiltonian was investigated in Ref.~[\onlinecite{biolatti02}], but there the off-diagonal interaction terms
($V_{\rm F}$) were neglected.

\begin{figure*}
\vspace{2cm}
\centerline {\hspace{-3cm}\includegraphics[bb=0 0 604 224, scale=0.5]{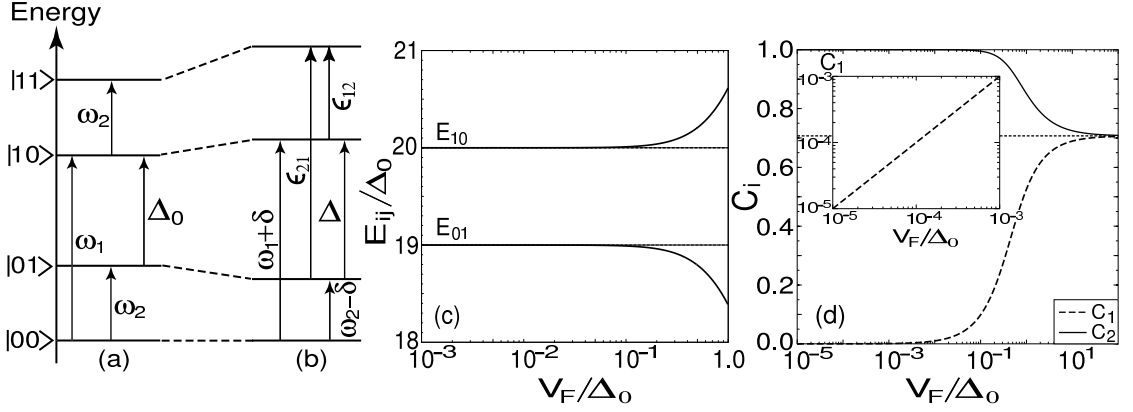}}\vspace{0cm}
\caption{A schematic diagram of the properties of the model Hamiltonian, Eq.~\ref{eq:Hdot}:
(a) energy levels in the absence of qubit (interdot) interactions; (b) energy levels in the presence of qubit
interactions for two dots (I and II) of different excitation frequencies.
$\epsilon_{\rm 12} = \omega_2 + V_{\rm XX} -\delta$,
$\epsilon_{\rm 21} = \omega_1 + V_{\rm XX} +\delta$, and $\delta\equiv V_{\rm F}^2/\Delta_0$,
where $V_{\rm F}$ and $V_{\rm XX}$ represent the strength of the F\"orster and direct Coulomb binding interactions
respectively.
In this case $V_{\rm F}\ll\Delta_0$. (c) Eigenenergies $E_{01}$, and $E_{10}$ corresponding to the qubit eigenstates
$\ket{\Psi_{01}}$, and $\ket{\Psi_{10}}$ as a function of the ratio $V_{\rm F}/\Delta_0$ for $\w_1/\Delta_0\equiv 20$.
For comparison, the dashed lines show the energies when $V_{\rm F}/\Delta_0=0$.
(d) The eigenstate coefficients $c_i$ as a function of the strength  $V_{\rm F}/\Delta_0$.
These coefficents show the
departure from the basis states $\ket{01}$, and $\ket{10}$ followed by the eigenstates
$\ket{\Psi_{01}}= c_1\ket{10}+c_2\ket{01}$ and $\ket{\Psi_{10}}=-c_1\ket{01}+c_2\ket{10}$.}
\label{QC}
\end{figure*}

The eigenenergies and eigenstates of the interacting  qubit system are
\beqa
\hspace{-0.15cm}
\left.
\begin{array}{cll}
\nonumber
E_{00}=&\w_0,                                                       &  \ket{\Psi_{00}}=\ket{00}; \\
E_{01}=&\w_0+\w_1-\frac{\Delta_0}{2}(1+A),  & \ket{\Psi_{01}}= c_1\ket{10}+c_2\ket{01}; \\
E_{10}=&\w_0+\w_1-\frac{\Delta_0}{2}(1-A),  & \ket{\Psi_{10}}=-c_1\ket{01}+c_2\ket{10}; \\
E_{11}=&\w_0+\w_1+\w_2+V_{\rm XX},          &\ket{\Psi_{11}}=\ket{11}, \\
\end{array}
\right. 
\eeqa
\vspace{-0.8cm}
\begin{equation}
\label{eigen}
\end{equation}
where $A=\sqrt{1+4(V_{\rm F}/\Delta_0)^2}$, $c_1 = \sqrt{(A-1)/2A}$ ($\approx V_{\rm F}/\Delta_0$ for
$V_{\rm F}/\Delta_0\ll 1$) and $c_2 = \sqrt{(A+1)/2A}$~.
The eigenenergies in the absence and presence of interdot interactions are displayed in Figs.~\ref{QC}(a), (b) and (c).
Fig.~\ref{QC}(a) shows the energy levels when the interactions are off; Fig.~\ref{QC}(b) shows these when the interactions
are on, but where $V_{\rm F} \ll \Delta_0$; Fig.~\ref{QC}(c) shows $E_{\rm 10}$ and $E_{\rm 01}$ as a function of the ratio
$V_{\rm F}/ \Delta_0$. Fig.~\ref{QC}(d) shows $c_1$ and $c_2$ as a function of  $V_{\rm F}/\Delta_0$.
These figures demonstrate that $V_{\rm F}$ causes a mixing of the states $\ket{01}$ and  $\ket{10}$
such that the eigenstates of the interacting system are not the same as the computational basis.
As we show below, this $V_{\rm F}$ coupling can be used for generating highly entangled states.\\

\begin{figure}
\vspace{3cm}
\centerline{\hspace{-1cm} \includegraphics[bb=0 0 479 339, scale=0.3]{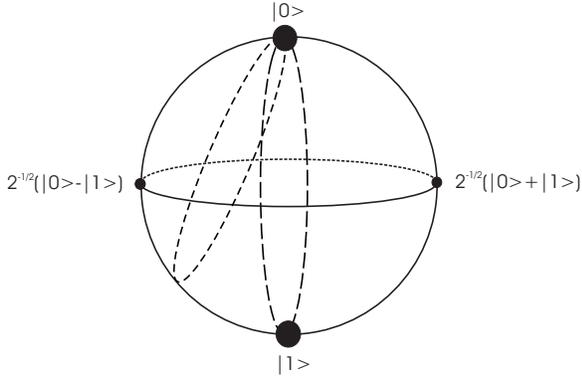}}\vspace{-2cm}
\caption{Bloch sphere representation of a single qubit. In order to perform an arbitrary single
qubit rotation in the exciton system, it is necessary to have the ability to induce two different Rabi
oscillations. Example trajectories for two such oscillations are shown by dashed lines.}
\label{singlebloch}
\end{figure}

Single qubit operations can be achieved by inducing
Rabi oscillations in the excitonic system~(e.g., see Refs.~\onlinecite{kama01} and~\onlinecite{borri02}).
If we take a Bloch sphere representation of a qubit, where the state $\ket{0}$ is represented by a
unit vector from the origin to the north pole of the Bloch sphere and the state $\ket{1}$ by a unit
vector to the south pole, then the qubit state
$\ket{\psi}=\exp(i\lambda)(\cos(\theta/2)\ket{0}+\exp(i\varphi)\sin(\theta/2)\ket{1})$,
where $\lambda$ is a global phase, is defined by the unit vector
$(\cos\varphi\sin\theta,\sin\varphi\sin\theta,\cos\theta)$.
Depending of the values of $\theta$ and $\varphi$, this vector may point to any point on the surface of the sphere, and for
universal quantum computation it must be possible to move the vector between any two of these points;
this defines an arbitrary single qubit rotation.
In our QD system, this control can be achieved by using laser pulses to induce two distinct Rabi oscillations
(see Fig.~\ref{singlebloch}).
The energy and length of such pulses must take into account structural
factors like the dot confinement energies and transition dipole moments.~\cite{borri02,schelpe02}
It is also essential that the exciton states have long enough
decoherence times that control over the phase $\varphi$ is possible.~\cite{reina02,nazir03}
For example, self-assembled semiconductor  (e.g. InGaAs/GaAs) quantum dots
could be advantageous for qubit manipulations since they exhibit
large dipole moments and long dephasing times.~\cite{kama01, borri02}
We shall return to the role of the QD material composition parameters for QC below.

\begin{figure}
\vspace{6cm}
\centerline{\hspace{-2cm} \includegraphics[bb=0 0 355 222,scale=0.45]{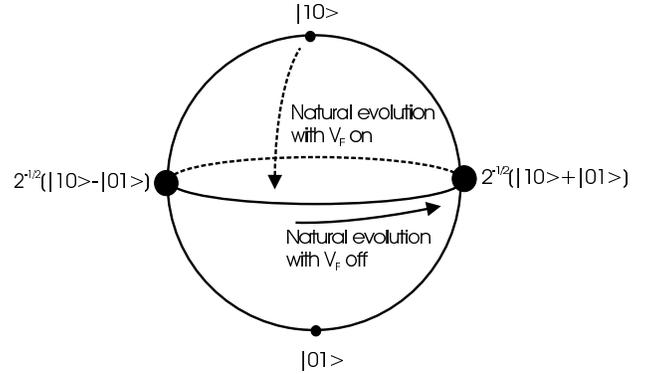}}\vspace{-4.7cm}
\caption{Illustration of how to create an entangled state by manipulating the off-diagonal F\"orster
coupling between two nanostructures. When the interaction is on and much greater than $\Delta_0$, the eigenstates
of the system
correspond to the large dots on the equator of the Bloch sphere. Thus after selective excitation of state $\ket{10}$
the system will naturally evolve to the equator. The interaction may then be suppressed by means of
an applied field, and the state remains maximally entangled as the eigenstates become the computational basis
states.}
\label{forsterbloch}
\end{figure}

The $V_{\rm XX}$ and $V_{\rm F}$
interactions lead to three possible ways of achieving quantum entanglement.
First, if the ratio $V_{\rm F}/\Delta_0 \gg 1$, the eigenstates of the system are approximately
$\ket{00}, \raiz(\ket{10} - \ket{01}), \raiz(\ket{10} + \ket{01})$ and $\ket{11}$.
We now further assume that the ratio $V_{\rm F}/\Delta_0$ can be controlled, by means of applying an
electric field to either change $V_{\rm F}$ directly, or to increase $\Delta_0$ by means of the Stark shift (we shall
discuss both of these effects in detail in Section~\ref{VF}). Then, we initially prepare the system in a state where
$V_{\rm F}/\Delta_0\ll 1$ and we selectively
excite QD I and create $\ket{10}$. Now, when the F\"orster interaction is turned on, the system will naturally evolve
sequentially into the following states:
$\ket{10} \mapsto \raiz(\ket{10} + i\ket{01}) \mapsto \ket{01} \mapsto \raiz(\ket{10} - i\ket{01}) \mapsto \ket{10}$
(see Fig.~\ref{forsterbloch}). This evolution could be then stopped when the system is in a maximally entangled state by
applying an electric field to suppress the F\"orster coupling once more, an effect which we shall again explore in detail
in Section~\ref{VF}.

Second, if the system does not have a strong F\"orster coupling, i.e. $V_{\rm F}/\Delta_0\ll 1$ all the time, the computational basis states are essentially the eigenstates of the system.
Then the $V_{\rm XX}$ coupling term implies that the resonant frequency for transitions between the
basis states $\ket{0}$ and $\ket{1}$ of
one qubit \emph{depends} on the state of the neighbouring qubit. This means that it is possible to construct
a \textsc{cnot} gate in this system. Such a gate flips the
target qubit $k$ if the control qubit $j$ is in the state $\ket{1}$ and acts
trivially otherwise:
$\textsc{cnot}_{jk}(\ket{m}_j\ket{n}_k)\mapsto\ket{m}_j\ket{m \oplus n}_k$,
where $m,n\in \{0,1\}$, and $\oplus$ denotes addition modulo 2 or XOR
operation.
By referring to Fig.~\ref{QC}(b), we can see that the logic operation
$\textsc{cnot}_{12}(\ket{1}_1\ket{0}_2)\mapsto \ket{1}_1\ket{1}_2$ can be achieved by illuminating the qubit system in
the state $\ket{10}$ with a $\pi$-pulse of energy  $\epsilon_{12} = \w_2+V_{\rm XX}-\delta$ (a pulse which we label
$\pi_{\epsilon_{12}}$).
Conversely, if the role of the control qubit is to be performed by the second qubit, the gate operation
$\textsc{cnot}_{21}(\ket{0}_1\ket{1}_2)\mapsto \ket{1}_1\ket{1}_2$ can be realized via the application of a
$\pi$-pulse of energy  $\epsilon_{21} = \w_1+V_{\rm XX}+\delta$
(a $\pi_{\epsilon_{21}}$ pulse) to the system state $\ket{01}$.
Crucially, the energy difference between these two
$\pi$-pulses, and hence the energy selectivity of the logic gate, is determined by
$\Delta_0$, $V_{\rm XX}$ and $V_{\rm F}$.
From Fig.~\ref{QC}(b) we can also see how to use the $\textsc{cnot}$ gate to create maximally entangled states. For example,
if we start in the ground state and first apply a $\pi/2$ or $3\pi/2$ pulse at energy
$\w_1$, we create the states $\raiz (\ket{00}\pm\ket{10})$; if we now apply a $\pi_{\epsilon_{12}}$ pulse,
we generate the maximally entangled states $\raiz (\ket{00}\pm\ket{11})$.

Third, the interaction with the laser field does not necessarily have to
be the entangling mechanism in the case where $V_{\rm F}/\Delta_0\ll 1$. If each of two neighbouring single qubits are
prepared in the superposition state $\raiz (\ket{0}+\ket{1})$  (i.e. a Hadamard transform is applied to both of them),
making the state $\half (\ket{00}+\ket{01}+\ket{10}+\ket{11})$, this will then naturally evolve into entangled states
under the action of $V_{\rm XX}$ alone.

Although we have discussed our Hamiltonian specifically in the context of a dot molecule, these ideas are valid for any nanostructure with corresponding diagonal and off-diagonal interactions
where an energy selective excitation is possible.
However, we  focus on a quantum dot implementation, and in the next sections we shall analyze in detail the inter- and
intra-dot interaction terms and show how they can be tailored as a function of the interdot distance, dot sizes, and
material composition.

\section{The Model System}
\label{model}
\begin{figure}
\vspace{2cm}
\centerline {\includegraphics[bb=0 0 349 541,scale=0.3]{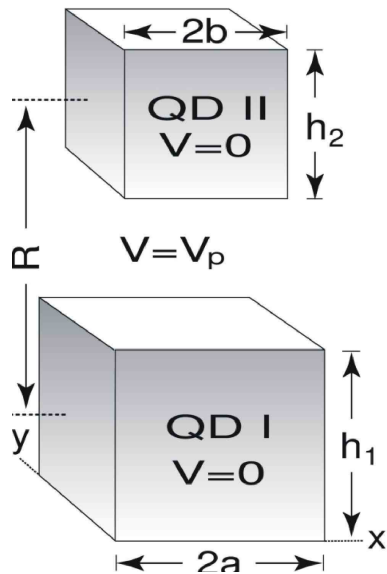}}\vspace{0cm}
\caption{Schematic diagram of the cuboidal dot model. The cuboids have base sides of length $2a$ and $2b$ and
heights of $h_1$ and $h_2$ respectively.
Their centres are separated by a distance $R$. The potential inside the cuboids is set to zero, and that outside
them is determined by the band offsets of the conduction and valence bands within the heterostructure.}
\label{dotspic}
\end{figure}

Different dot geometries (e.g. spherical, pyramidal or cuboidal shaped dots)
can be used to implement the logic gates and quantum entanglement schemes discussed above, and in this article
we choose dots of a
square-based cuboidal shape (for another possible geometry see Ref.~\onlinecite{nazir03}).
We assume that the potential energy $V$ of both electrons and holes
increases abruptly at the
cuboid boundaries where the semiconductor bandgap changes, and that $V=0$ inside the cuboids (see Fig.~\ref{dotspic}).
The confinement potential is determined by the band offsets for the electrons and the holes.
This type of square well potential has the advantage of
describing both a well defined dot size in all three dimensions and of
bound and unbound solutions in each direction (this is in contrast with, for example, the
parabolic potential considered in Ref.~\onlinecite{biolatti02}).

\begin{figure}
\vspace{3cm}
\centerline{\hspace{-2.5cm} \includegraphics[bb=0 0 575 302,scale=0.3]{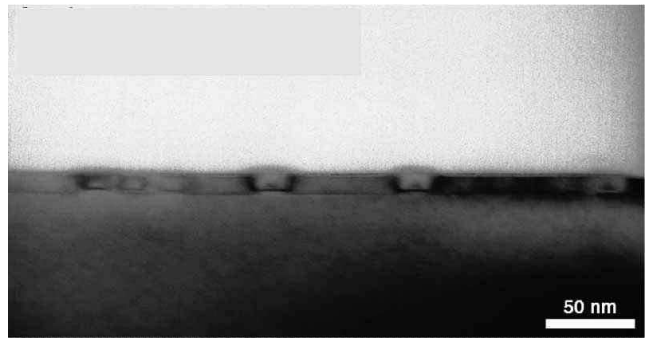}}\vspace{0cm}
\caption{Transmission electron micrograph of two layers of quantum dots grown by the Stranski-Krastanow method.
The dots are made of InAs and the
encapsulating material is GaAs; note that the dots in the second layer nucleate preferentially above the dots in the
first layer. (Figure courtesy of C.~Lang, C.~Marsh and D.~Cockayne (Oxford Materials);
sample courtesy of M.~Hopkinson and P.~Houston (Sheffield University).)}
\label{SKdots}
\end{figure}

Our model captures the essential properties of quantum dots which are grown by the Stranski-Krastanow (SK)
method.~\cite{eaglesham90} Such structures show a degree of self-organization~\cite{xie95}
which is ideally suited for the manufacture of prototype quantum devices,
and in the realization of the elementary quantum logic gates discussed in this paper.
The SK dot growth proceeds through the evaporation of a layer of dot material on to a previously grown substrate:
dots form spontaneously
due to the competing energy contributions of dot surface area, dot volume and strain during the growth process.
After the formation of dots and subsequent overgrowth of the substrate material, a second layer of dots may be grown;
these nucleate preferentially above the dots in the first layer due to the uneven strain field at the
surface.~\cite{tersoff96}
An example of two dots grown in this way is shown in Fig.~\ref{SKdots} -- and it can be seen there that
two vertically stacked dots have been grown with a
controlled spacing; dots with such characteristics may be well suited for performing the logic operations described above.
As can be seen in Fig.~\ref{SKdots}, the dots tend to have smaller dimensions in the growth
direction than in the perpendicular directions, and the upper dot of the pair tends to be of a slightly larger size,
thus allowing for an appropriate identification of the excitation frequencies $\w_1$ and $\w_2$ required in our model.
Qubit scalability is available via the SK growth procedure since several layers of dots have been shown to grow
in stacks.~\cite{solomon96}

Our computational Hilbert
space requires up to one exciton per dot, and we must therefore calculate both the
single particle energies, which are determined by the potential profile of the dots, and the
two- and four-particle interactions which are determined by the strength of the Coulomb interaction between
them. We look at these different quantities separately in this section.

\subsection{Single Particle States}
\label{spstates}
There are a variety of methods for finding the solutions of the Schr\"odinger equation for electrons or holes in
a quantum dot. These include a full pseudopotential calculation,~\cite{franceschetti97, franceschetti,williamson00}
finite element analysis,~\cite{johnson98} plane wave expansion~\cite{cusack96}
and the use of finite differences.~\cite{pryor} We employ a similar strategy to that in
Refs.~\onlinecite{califano} and \onlinecite{gang97}, where the Schr\"odinger equation is expanded in a set of analytical
basis functions which are the exact solutions of a potential which is close to the one under investigation.
This method has the advantage that the state solutions can be stored in a vector in Hilbert space rather than
as a wavefunction amplitude at each of a very large number of different spatial points; time evolution of the
quantum states is also easier to simulate when the state is represented by a vector
and we shall extend our work to this area elsewhere. Furthermore, a vector representation can allow
more physical insight since the basis functions themselves have a known physical interpretation.

Our first step is to express the wavefunctions for single particles in the envelope
function approximation as:~\cite{harrison00, bastard81}
\begin{equation}
\psi_p({\bf r}) = \phi_p({\bf r}) U_p({\bf r})
\label{envelope}
\end{equation}
where $\phi_p({\bf r})$ is an envelope function describing the changing wavefunction amplitude of confined states
for particle type $p$
over the dot region, and $U_p({\bf r})$ is the Bloch function which has the periodicity of the atomic lattice. In
the effective mass approximation, the
envelope functions are solutions of the following single particle Schr\"odinger equation:
\begin{equation}
\label{eq:spschr}
\left[-\frac{\hbar^2}{2}\nabla\left(\frac{1}{m^\ast_{p}({\bf r})}\right)\nabla +
V_{p}({\bf r})\right]\phi_p^i({\bf r}) = E^i_{p} \phi_p^i({\bf r})
\end{equation}
where $V_p$ is the confinement potential, which is displayed in Fig.~\ref{dotspic} and $m^\ast_p$ is the
effective mass of the particle $p$.
These solutions may be obtained by expanding the Hamiltonian in a
set of envelope basis functions of the form $\Xi({\bf r}) = \xi_x(x)\xi_y(y)\xi_z(z)$,
where the $\xi_i(i)$ are
the solutions
of a one dimensional square well potential with the appropriate effective masses.~\cite{califano, gang97}
Both bound and unbound states must be used in the expansion in order to obtain convergent solutions: the forms of these
are discussed in Appendix A.

\begin{figure}
\vspace{4cm}
\centerline {\hspace{-1cm}\includegraphics[bb=0 0 175 305,scale=0.8]{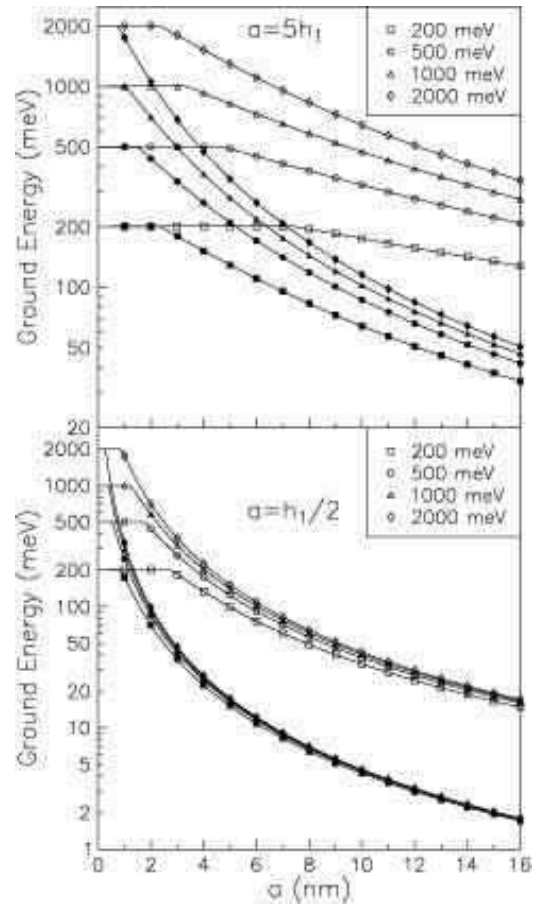}}\vspace{0cm}
\caption{The single particle ground state energy as a function of dots size for a quantum cube (lower graph) and a quantum
cuboid (upper graph). We use
the envelope function and effective mass approximations. The filled symbols correspond to $m_p^\ast=0.6m_0$ which
is typical of heavy holes, and the open symbols correspond to $m_p^\ast=0.06m_0$ which
is typical of electrons. These values will be used for electrons and holes throughout this paper.
The different symbols represent different values for the confinement potential (see legend).}
\label{onepart}
\end{figure}

There are two important things to mention about the direct expansion technique. First, the basis functions
we have described above do not in general form an orthogonal set if the bound state solutions have not
decayed to zero at the artificial infinite barrier which is used to generate the unbound states. In practice
this is rarely a problem, but anyway
is circumvented by using a modified basis set which is orthogonal and whose components are linear
combinations of the original basis functions. This modified basis set spans the same Hilbert space
as the original set. The method we employ to find this set is canonical orthogonalization~\cite{szabo89} which relies on
direct diagonalization of the matrix whose elements are the overlap integrals of the basis functions. The
Hamiltonian may then be expressed in this new basis as an Hermitian matrix, and solutions are found again by
direct diagonalization (we use the NAG diagonalization algorithm in our simulations).
The second point
is that the basis set must necessarily be truncated; hence the eigenenergies of the
solutions we obtain are really upper limits on the true eigenenergies of the coupled dot system
(we employ the Rayleigh-Ritz variational method~\cite{cohen}). In practice we can
increase the number of basis functions until a sufficiently accurate solution is obtained. We shall mention any important
points relating to solution convergence and approximations at the appropriate places later in the paper.

Results of a simple single particle calculation are displayed in Fig.~\ref{onepart}, where the ground state energy
of a two particles in QD I (of masses $0.6m_0$ and $0.06m_0$, where $m_0$ is the free electron mass) is shown as a
function of
the dot size for two dot geometries. The two geometries correspond to a cubic shape ($a=h_1/2$) and a
flatter cuboidal shape ($a=5h_1$), which is more typical of SK dots (see Fig.~\ref{SKdots}). As would be expected,
the ground state energy decreases for a larger dot size and is smaller for heavier particles.
The cube shaped dots have ground states with smaller energies than the corresponding cuboidal dot states
since the cuboids have one
smaller
dimension which increses the kinetic energy of the wavefunction.
All of the curves have a kink, and for dot sizes below this the ground energy saturates to the value of the
confinement potential. This is a consequence of including unbound states in the calculation: once the dots are small
enough that
the confined state energy has a larger energy than the confinement potential, the ground state becomes unbound and
there is hardly any dependence on dot size as it is further reduced. When the ground state is a bound
state, the ground energy is always larger in the case of the cuboidal shape, for a given value of $a$:
this is a consequence of the smaller height dimension of the cuboid which increases the kinetic energy of the
particle.
Bound ground states are always very closely approximated by the basis function corresponding to the ground state of the
one dimensional well in all three dimensions: in the case of cubic shaped dots, the amplitude of this state is always
greater than 0.999; for
the cuboidal dot it is always greater than 0.99. We shall use the approximation that the ground state is {\it exactly} this
basis function later in the paper.

\subsection{Coulomb Interactions: General Methodology}
\label{coulomb}
In this section we present our general methodology for calculating the Coulomb interaction
matrix elements between electrons and holes in quantum dots. This will be important later for
calculations of the intra- and interdot matrix elements in our model system.

First, consider an initial wavefunction of an $N$-electron system which represents a
single exciton state of a quantum dot:~\cite{franceschetti}
\beqa
\Psi_{\rm I}
& = & {\cal{A}}\Big[\psi_1({\bf r_\textrm{1}, \sigma_\textrm{1}}), \psi_2({\bf r_\textrm{2}, \sigma_\textrm{2}}),...,  \nonumber \\
&&\hspace{0.4cm}\psi_s^\prime({\bf r_\textrm{s}, \sigma_\textrm{s}}),...,\psi_t({\bf r_\textrm{t}, \sigma_\textrm{t}}),...,\psi_N({\bf r_\textrm{N}, \sigma_\textrm{N}})\Big],
\label{Xwave1}
\eeqa
where the $\cal{A}$ indicates overall antisymmetry (i.e., the wavefunction takes a Slater determinant form), the $\sigma_i$ represent the spin state of each electron, and the $\psi_i$
are single particle wavefunctions;
we have labelled the state $s$ with a prime symbol to indicate that it lies in the conduction band, whereas all of the other
states are in the valence band.

Next, we assume a final state which is a different single exciton:
\beqa
\Psi_{\rm F}
&=& {\cal{A}}\Big[\psi_1({\bf r_\textrm{1}, \sigma_\textrm{1}}), \psi_2({\bf r_\textrm{2}, \sigma_\textrm{2}}),...,   \nonumber \\
&&\hspace{0.4cm}\psi_s({\bf r_\textrm{s}, \sigma_\textrm{s}}),...,\psi_t^\prime({\bf r_\textrm{t}, \sigma_\textrm{t}}),...,\psi_N({\bf r_\textrm{N}, \sigma_\textrm{N}})\Big].
\label{Xwave2}
\eeqa
The Coulomb matrix element between these two states is given by:
\beq
M_{\rm IF}^{Coul} =
\bra{\Psi_{\rm F}}\sum_{ij, i<j}\frac{e^2}{4\pi\epsilon_0\epsilon_r({\bf r_{\textrm{ij}}})|{\bf r_{\textrm{ij}}}|}\ket{\Psi_{\rm I}},
\eeq
where $\epsilon_0$ is the permittivity of vacuum, $\epsilon_r ({\bf r})$ is the relative permittivity of the medium (and
therefore describes polarization screening), and ${\bf r_{\textrm{ij}}} = {\bf r_\textrm{i}}-{\bf r_\textrm{j}}$.
The only non-zero terms in the above expansion are those involving both ${\bf r_\textrm{s}}$ and ${\bf r_\textrm{t}}$, since
the ground and excited states of each single particle are orthogonal to one another. Hence, we obtain:
\beq
M_{\rm IF}^{Coul} = \bra{\Psi_{\rm F}}\frac{e^2}{4\pi\epsilon_0\epsilon_r({\bf r_{\textrm{st}}})|{\bf r_{\textrm{st}}}|}\ket{\Psi_{\rm I}}.
\eeq
Owing to the antisymmetric nature of the wavefunctions, this matrix element has contributions from a direct term and
an exchange term. Both terms take the form of an integral over ${\bf r_\textrm{s}}$ and ${\bf r_\textrm{t}}$. In spite of the fact
that it arises from the exchanged form of the wavefunction, the direct term is conventionally
written as:
\beqa
\nonumber
M_{\rm IF}^{J} = C\!\!\int\!\!\!\!\int\! \psi_t^{\prime\ast}({\bf r_\textrm{s}}) \psi_s^\prime({\bf r_\textrm{s}})
\frac{1}{\epsilon_r({\bf r_{\textrm{st}}})|{\bf r_{\textrm{st}}}|}
\psi_s^\ast({\bf r_\textrm{t}}) \psi_t({\bf r_\textrm{t}}) d{\bf r_\textrm{s}} d{\bf r_\textrm{t}},
\eeqa
\vspace{-0.7cm}
\begin{equation}
\label{J}
\end{equation}
and the exchange term is
\beqa
\nonumber
M_{\rm IF}^{K} = \pm C\!\!\int\!\!\!\!\int\! \psi_s^\ast({\bf r_\textrm{s}})\psi_s^\prime({\bf r_\textrm{s}})
\frac{1}{\epsilon_r({\bf r_{\textrm{st}}})|{\bf r_{\textrm{st}}}|}\psi_t({\bf r_\textrm{t}})
\psi_t^{\prime\ast}({\bf r_\textrm{t}})d{\bf r_\textrm{s}} d{\bf r_\textrm{t}}
\eeqa
\vspace{-0.7cm}
\begin{equation}
\label{K}
\end{equation}
where we have introduced the constant $C\equiv \frac{e^2}{4\pi\epsilon_0}$.
The sign of this exchange term is determined by the spin of the two particles: spin triplet ($S=1$) states have positive
$M_{\rm IF}^{K}$ elements, whereas spin singlet ($S=0$) states have negative $M_{\rm IF}^{K}$ values. We have removed the
spin variables from the single particle wavefunctions since these are not important for spatial integrals, once the sign
for $M_{\rm IF}^{K}$ has been determined.
It is natural now to switch to a hole description of the many body wavefunctions
introduced above: the matrix elements obtained by the above procedure would be identical if we labelled the states
$s$ and $t$ as holes and simply expressed each wavefunction as a product of the promoted electron states and the
left behind hole states. This allows the other parts of the wavefunctions to be left out when calculating matrix elements
involving these specific electronic states and so makes calculations easier; it also explains the convention on labelling
direct and exchange terms. We shall henceforth use such a description.

We proceed further by noting that the two integrals, Eq.~\ref{J} and Eq.~\ref{K}, are both of the form
\beq
I = \int \!\! \int \rho_s({\bf r_\textrm{s}}) f({\bf r_\textrm{s}- r_\textrm{t}})\rho_t({\bf r_\textrm{t}}) d{\bf r_\textrm{s}} d{\bf r_\textrm{t}} ,
\label{I1}
\eeq
and so we may use Fourier transforms to reduce the dimensionality of the integrand. Employing the convolution
theorem and Parseval's relation leads to
\beq
I = \frac{1}{(2\pi)^6}\int R_s({\bf K)} R_t({\bf K}) F({\bf K}) d{\bf K}  ,
\label{I2}
\eeq
where the Fourier transform of $\rho$ is denoted by $R$ and that of $f$ is denoted by $F$. We now make the calculation
more specific by first assuming that $\epsilon_r$ is independent of ${\bf r_{\textrm{st}}}$ (see Ref.~\onlinecite{franceschetti} for
a detailed discussion of the form of $\epsilon_r({\bf r_{\textrm{st}}})$),
and note that the Fourier transform of the Coulomb operator is given by
\beq
\label{Coulomb}
F({\bf K}) = \frac{4\pi}{K^2}\ .
\eeq
The $\rho$ functions are a product of two wavefunctions of the form of Eq.~\ref{envelope}, and their Fourier transform is
simplified by invoking the different lengthscales of the envelope function and the Bloch function. We may write, for
the $M_{\rm IF}^{K}$ integral (a completely analogous method can be carried out for the $M_{\rm IF}^{J}$ integrals):
\beqa
R_s({\bf K}) &=& \int \psi_s^\ast({\bf r_\textrm{s}}) \psi_s^\prime({\bf r_\textrm{s}}) \exp(i{\bf K \cdot r_\textrm{s}})
d{\bf r_\textrm{s}}  \nonumber\\
&=& V_{cell}\sum_{{\bf T}_i} \phi_{s,i}^\ast \phi_{s,i}^\prime \exp(i{\bf K \cdot T}_i)\times \nonumber\\
&&\int_{cell} U_s^\ast({\bf r_\textrm{s}}) U_s^\prime({\bf r_\textrm{s}})\exp(i{\bf K \cdot r_\textrm{s}})
d{\bf r_\textrm{s}}\ ,
\eeqa
where we have assumed that each envelope function takes a constant value, $\phi_i$, over each unit cell $i$ of the lattice,
that the translational lattice vector of cell $i$ is ${\bf T}_i$ and that the volume of the unit cell is $V_{cell}$.
We may express the ${\bf K}$ wavevector as
\beq
{\bf K = k+G}\ ,
\eeq
where ${\bf k}$ is a vector within the first Brillouin zone and ${\bf G}$ is a reciprocal lattice vector, and
we may further convert the sum to an integral to obtain
\beqa
\label{Rs}
R_s({\bf k, G}) &= &\int_{space} \phi_s^\ast({\bf r}) \phi_s^\prime({\bf r}) \exp(i{\bf k \cdot r}) d{\bf r}\times  \\
&& \int_{cell} U_s^\ast({\bf r_\textrm{s}}) U_s^\prime({\bf r_s})\exp(i{\bf (k+G) \cdot r_\textrm{s}}) d{\bf r_\textrm{s}}\
.  \nonumber
\eeqa
Thus $R_s$ is a product of Fourier transforms, one for the
envelope functions (which is independent of ${\bf G}$, so only needs to be calculated within the first Brillouin zone)
and the other for the Bloch functions. It is also obvious that an analogous expression exists for $R_t$.
The envelope function parts of the Fourier transform are analytical: the wavefunction takes either a sinusoidal or
exponential form depending on whether it exists within or outside the quantum dot. We do not write out these expressions
explicitly here however, since they are somewhat lengthy and tedious. The Bloch function part is also analytical for a
suitable choice of wavefunction, and to simplify the calculations here we use a Kronig-Penney model where the atomic
wavefunctions are assumed to take the form of the
solutions of an infinite square well potential of well width $2x$. Specifically, we assume that the
hole states we consider (at the top of the valence band) take the wavefunction
solution of this potential which has $p_z$ symmetry and the electron states we consider take the solution with
$s$ symmetry.
We expect this to be sufficient approximation for
elucidating the general properties of the system, though a more refined calculation would be required to obtain
more accurate estimates of the various quantities we calculate.

By inserting Eq.~\ref{Rs} (and the analogous expression for $R_t$) and Eq.~\ref{Coulomb} into Eq.~\ref{K}
we obtain an expression for $M_{\rm IF}^{K}$. The expression has an integrand which is analytical
but the integration over
three-dimensional ${\bf K}$ space must be carried out numerically. This is done by employing a NAG
library routine for multi-dimensional adaptive quadrature.
If the wavefunction labels are swapped around, an analogous method for calculating $M_{\rm IF}^{J}$ can be carried out
(see Eq.~\ref{J}).
We shall show the results of such calculations for various different cases in the next sections.

\section{Intra-Dot coupling}
\label{intra}
In this section, we describe the predictions of the above model when it is applied specifically to the
calculation of the diagonal matrix element of the two ground state
basis functions representing an electron and a hole on the same dot.
In this case the states $s$ and $t$ are identical,
and hence the expression for $M_{\rm IF}^{J}$ reduces to the direct Coulomb interaction between the ground basis state
electron and the ground basis state hole (we call this $M_{\rm 00}^{J}$).
We saw in the previous section that the ground basis state is a good
approximation to the true ground state of the system when only the single particle contributions to the
Hamiltonian are taken into account. Thus this matrix element is a first order correction to the energy due to the
Coulomb force between the two particles.
Furthermore, the expression for $M_{\rm IF}^{K}$ reduces to the spin splitting between singlet and triplet
exciton states in this first order approximation.

\begin{figure}
\vspace{4cm}
\centerline {\hspace{-4cm}\includegraphics[bb=0 0 167 140,scale=0.9]{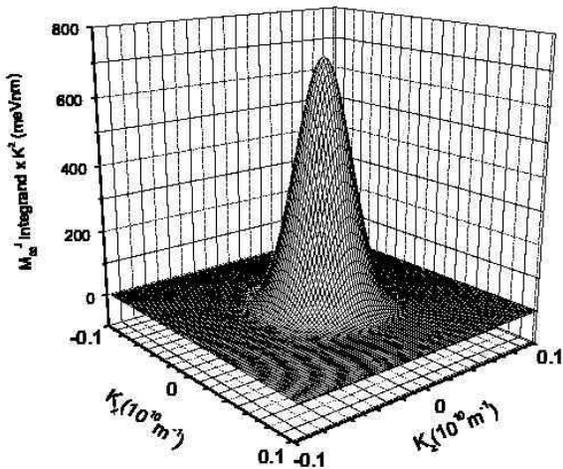}}\vspace{0cm}
\caption{The intra-dot direct Coulomb interaction strength integrand, plotted in the $K_x-K_z$ plane of
reciprocal space, and around the ${\bf K} = 0$ point. We have calculated the integrand for dot I
and used the cubic geometry ($a=h_1/2=10$~nm), and assumed that $V_e = V_h = 500$~meV.}
\label{Jpeak}
\end{figure}

Let us first consider the direct Coulomb interaction in dot I, which has basal side length $a$, and simplify things by
calculating for a cubic shape (i.e., we set $2a=h_1$). The results may then be directly
carried over to dot II, with $a \rightarrow b$ and $2b=h_2$.
As we described earlier, it is necessary to evaluate the integral
of Eq.~\ref{J} by first transforming to reciprocal space and then integrating over ${\bf K}$ space. The resultant integrand
has peaks at each reciprocal lattice point (where ${\bf k = 0}$ and so the envelope function part of the integrand has a
maximum). These peaks quickly die away over a lengthscale $\sim 1/a$ as would be expected for envelope functions representing
wavefunctions within QDs of side length $a$. However, it turns out that only the central (${\bf G = 0}$) peak is important,
since the other peaks contribute much less to the total integral (this is caused by both the ${\bf k}$ dependence of the
atomic contribution to the integrand
and by the $1/k^2$ dependence of the Coulomb interaction part). The central peak is displayed in Fig.~\ref{Jpeak}, as a
function of $K_x$ and $K_y$ ($K_z=0$).

\begin{figure}
\vspace{4cm}
\centerline {\hspace{-2cm}\includegraphics[bb=0 0 169 164,scale=0.9]{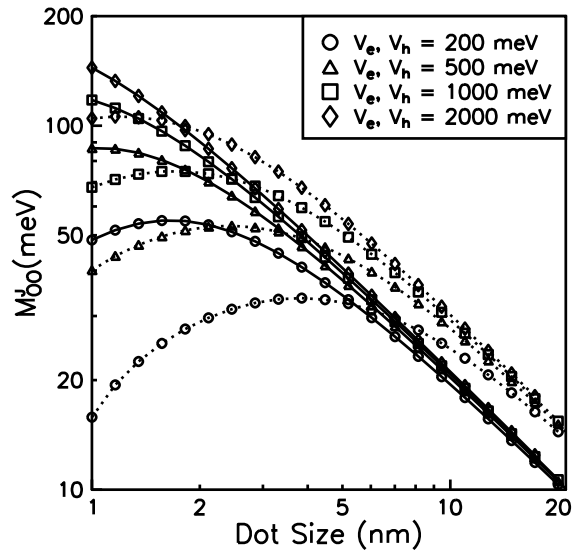}}\vspace{0cm}
\caption{The intra-dot direct Coulomb interaction strength, $M_{00}^{J}$ as a function of
dot size and confinement potential. The solid lines are for a cubic geometry ($a=h_1/2$),
and the dotted lines are for a cuboid ($a=5h_1$).}
\label{intraJ}
\end{figure}

By numerically integrating the central peak for a range of dot sizes and confinement potentials,
we can obtain a plot of the dependence of $M_{\rm 00}^{J}$ on these parameters
(this is shown in Fig.~\ref{intraJ} for both the cubic geometry and for a flat cuboid, in which $a=5h_1$).
As would be expected, the interaction decreases as the size of
the QD increases (and so the electron and hole are not forced to be so close together). It is interesting to
look at what happens at shorter distances when the confinement potential changes; a larger confinement potential
causes a larger Coulomb binding energy. This result is expected since the wavefunction of both the electron and
the hole is contracted when the confinement potential is larger -- and the resultant closer proximity of the
two wavefunctions causes a larger Coulomb interaction.
As would be expected intuitively, at very large dot sizes the size of the Coulomb interaction scales like $1/a$.
At very small dot sizes, the direct Coulomb interaction does not follow these simple rules: for a weak enough confinement
there is a peak in the
energy and at small values of $a$ it {\it decreases}. This can be understood by thinking about the shapes of the
wavefunctions in this region. When the well width is small, the curvature of the wavefunction is necessarily rather high,
and so the kinetic energy is large. In order to compensate for this, the wavefunction spreads out into
the barriers at the cost of some potential energy, if the barriers are not too high
(this energy cost is balanced by the saving in kinetic energy). Thus
the wavefunction has a larger size than would na\"{\i}vely be expected from the dot size, and the Coulomb binding energy
decreases. The peak in the Coulomb potential occurs for a larger value of $a$ in the cases of cuboidal geometry: this is
simply because of the shorter height dimension in this case, which means that the wavefunction spread effect discussed above
remains significant at larger values of $a$. At still larger values of $a$, the Coulomb interaction is larger for the
cuboidal geometry than it is for the cube: again this is because the cuboid has one smaller dimension, which means that the
electron and hole are forced to be closer together in the cuboidal case.

\begin{figure}
\vspace{4cm}
\centerline {\hspace{-4cm}\includegraphics[bb=0 0 166 140,scale=0.9]{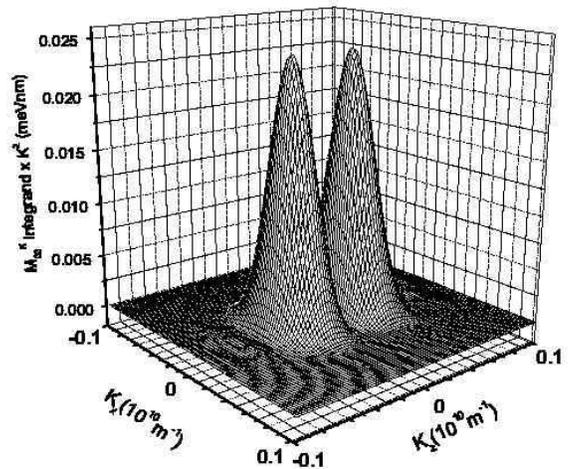}}\vspace{0cm}
\caption{The intra-dot exchange Coulomb interaction strength integrand, plotted in the $K_x-K_z$ plane of
reciprocal space, and around the ${\bf K} = 0$ point. Note that it has value zero at ${\bf K}=0$ and is asymmetric in
$x$ and $z$ due to the choice of the $p_z$ Kronig-Penney state for the holes. We have calculated the integrand for dot I
and used the cubic geometry ($a=h_1/2=10$~nm) and assumed that $V_e = V_h = 500$~meV.}
\label{Kpeak}
\end{figure}

\begin{figure}
\vspace{3cm}
\centerline{\hspace{-2cm}\includegraphics[bb=0 0 174 162,scale=0.9]{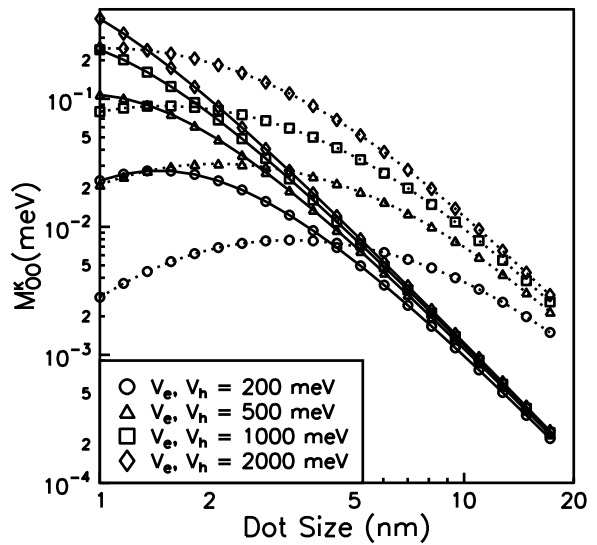}}\vspace{0cm}
\caption{The intra-dot exchange Coulomb interaction strength, $M_{00}^{K}$ as a function of
dot size and confinement potential. The solid lines are for a cubic geometry ($a=h_1/2$),
and the dotted lines are for a cuboid ($a=5h_1$).}
\label{intraK}
\end{figure}

We next look at the value of the exchange coupling between the two ground state basis functions (i.e. the spin
singlet-triplet splitting, $M_{\rm 00}^{K}$, correct to first order).
The relevant $K$-dependent integrand takes a somewhat different form
in this case. The central peak (around ${\bf G=0}$) is displayed in Fig.~\ref{Kpeak}, where it can be seen that the function
has a zero at $K=0$; this is expected since the electron and hole wavefunctions have opposite parity. The suppression
at $K=0$ means that, this time, the regions around other reciprocal lattice points have to be included in the numerical
integration. The resultant dependence of $M_{\rm 00}^{K}$ on dot size and
confinement potential is displayed in Fig.~\ref{intraK}.

By reference to Fig.~\ref{intraK} we see that the exchange splitting is several orders of magnitude smaller than the
direct Coulomb term, though it follows the same trends of increasing in value with smaller dot sizes and larger confinement
potentials. These effects
can be understood as follows: the exchange splitting is essentially a consequence of Pauli's exclusion principle which
states that particles in the same quantum state cannot exist together at the same spatial point. Thus electrons and holes
which are in a triplet spin state (and so have indistinguishable spin properties) necessarily `avoid' each other, thus
reducing the Coulomb attraction between them. This effect is expected to be more significant when the wavefunctions in the
absence of the Coulomb interaction overlap strongly -- that is when they are localized in one small region of space.
There is, in general,
a greater degree of localization of the wavefunctions when either the dots are smaller or when the confinement is stronger,
and hence the exchange splitting gets larger when these conditions are satisfied.
However, there is again one region of Fig.~\ref{intraK} where this general rule is not obeyed - that is when confinement is
relatively weak,
but where the dot size is small. Here, the exchange energy takes a down turn as the dot size gets smaller. This is caused by
increased wavefunction barrier penetration, which again means the effective wavefunction size gets {\it larger} rather than
smaller as might be expected from the simple intuitive picture described above. The comparison between the cubic and
cuboidal geometry show a similar trend to that discussed above for the case of the direct Coulomb interaction, and the
reasons
for this follow the same lines. The turnover occurs for a larger $a$ for the cuboidal dot due to the wavefunction spreading
effect which occurs for smaller spatial dimensions: this spreading causes a reduction in the exchange splitting. At larger
sizes, the exchange splitting is larger for the cuboidal shape due to the effect of the smaller height dimension of the
cuboid, which pushes electron and hole together and thus increases the exchange splitting.
At larger dot sizes, the interaction scales like $1/a^3$.

\section{Inter-Dot Interactions}
\label{inter}
In this section we shall discuss the {\it inter}-dot coupling terms which are due to the Coulomb operator introduced in
previous sections. These terms are crucial to the operation of a quantum device, since they may allow qubit-qubit
interactions to take place, which is an essential requirement for two (or more) qubit gates to be constructed.
There are two important types of interaction which may occur. The first type is called the F\"orster interaction, and is
described by an off-diagonal matrix element (in the computational basis) between two single exciton wavefunctions of the
type introduced in Eqs.~\ref{Xwave1}
and~\ref{Xwave2}, but where the two excitons are located on different quantum dots (this interaction is called $V_{\rm F}$
in Eq.~\ref{eq:Hdot}). The second interaction which is important for this scheme is the direct self Coulomb interaction
in a biexciton (double substitutional Slater determinant) wavefunction, where one exciton is located on each dot. This is
called $V_{\rm XX}$ in Eq.~\ref{eq:Hdot} and amounts to the Coulomb binding energy between two excitons located on
adjacent dots. We next  quantify both of these interaction terms and discuss their properties within the context of the
quantum computing implementation described in Sec.~\ref{heisenberg} .

\subsection{Off diagonal coupling: F\"orster interaction}
\label{VF}
\begin{figure}
\vspace{4cm}
\centerline {\hspace{-4cm}\includegraphics[bb=0 0 161 140,scale=0.9]{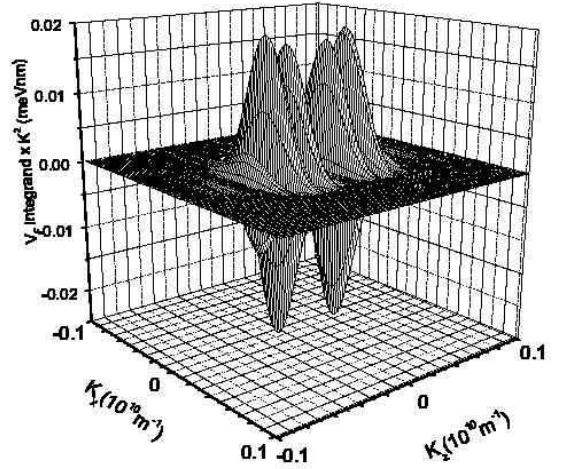}}\vspace{0cm}
\caption{The plot of the integrand in ${\bf K}$ space which leads to the F\"orster strength, as a function of
$K_x$ and $K_z$ ($K_y=0$). The extra modulation on the function (as compared with Figs.~\ref{Jpeak} and~\ref{Kpeak})
is caused by the interdot separation. The plot is for cubic dots with $a=b=10$~nm and $R=20$~nm, and we have taken
$V_e = V_h = 500$~meV.}
\label{fig:Forstplot}
\end{figure}

\begin{figure}
\vspace{3cm}
\centerline {\hspace{-2cm}\includegraphics[bb=0 0 171 164,scale=0.9]{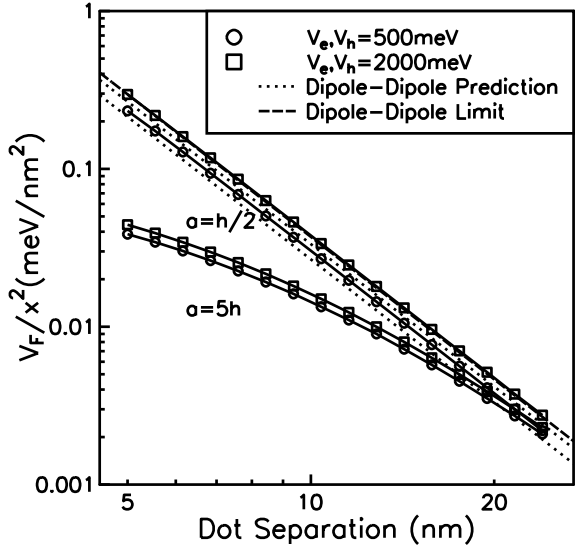}}\vspace{0cm}
\caption{Dependence of the F\"orster interaction strength on the interdot separation $R$.
The dots have equal sizes and results are shown for two shapes: (i) cubic, with  $a=h/2=2$~nm (upper two curves)
and (ii) cuboidal, with $a=5h=2$~nm (lower two curves).
The circles and squares represent the predictions of a full numerical
simulation for well depths of 500~meV and 2000~meV respectively
(the electron and hole wells are assumed to be of the same depth).
The dotted lines represent the predictions of the dipole-dipole model for the cubic shaped dots
and the dashed curve
represents the maximum coupling predicted for the dipole-dipole model, which corresponds to $O_{\rm I} = O_{\rm II} = 1.0$.
The dotted lines can be obtained by multiplying this maximum by the relevant values of $O_{\rm i}$ which can be obtained
from Fig.~\ref{fig:overlap}.}
\label{fig:forster}
\end{figure}

\begin{figure}
\vspace{3cm}
\centerline {\hspace{-2cm}\includegraphics[bb=0 0 171 164,scale=0.9]{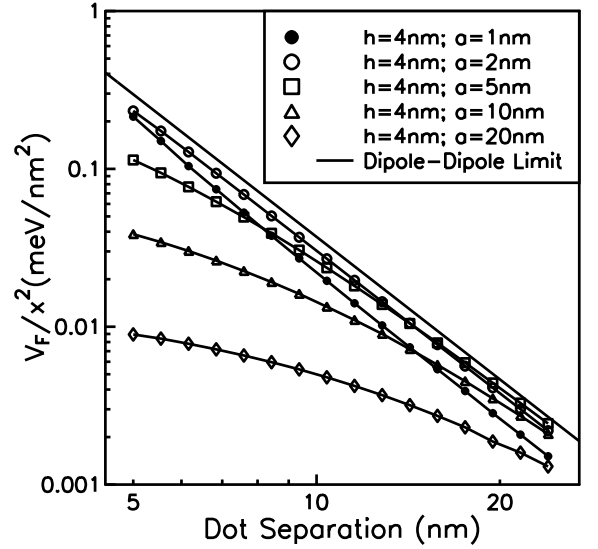}}\vspace{0cm}
\caption{Dependence of the F\"orster interaction strength on the shape of the dots. The dots are assumed to be identical,
but have a series of cuboidal shapes with different aspect ratios. The well depth is 500 meV
(the electron and hole wells are assumed to be of the same depth).
The solid curve
represents the maximum coupling predicted for the dipole-dipole model, which corresponds to $O_{\rm I} = O_{\rm II} = 1.0$.}
\label{fig:newforster}
\end{figure}

The F\"orster or Coulomb exchange interaction
can induce the transfer of an exciton from
one quantum dot to the other. This is a non-radiative energy transfer whereby an exciton is destroyed on one
dot and recreated on the other; it is an electrostatic interaction which proceeds via a short lived
virtual photon. F\"orster's original theory~\cite{forster48} showed that the interaction is dipole-dipole to lowest order; this theory was
subsequently elaborated by Dexter~\cite{dexter53} who derived higher order and exchange terms in studies of the sensitized
luminescence of solids. Here we extend this theory
to the case of the many-body exciton states of quantum dots. The off-diagonal nature of the interaction
causes the eigenstates of the Hamiltonian, Eq.~\ref{eq:Hdot},
to become linear combinations of the computational basis states $|10\rangle$ and $|01\rangle$.
As described in Section~\ref{heisenberg}, the degree of this mixing is crucial in determining how to generate
quantum entanglement in the quantum dot molecule.
The F\"orster coupling can be expressed as the matrix element of the direct Coulomb operator
between excitons located on each of the two dots:
\beqa
\label{Forster}
V_{\rm F} &=& C \int\!\!\int \psi_s({\bf r_\textrm{s}})\psi_s^\prime({\bf r_\textrm{s}})\times \\ \nonumber
&&\frac{1}{\epsilon_r({\bf R+r_\textrm{s}-r_\textrm{t}})|{\bf R+r_\textrm{s}-r_\textrm{t}}|}\,
\psi_t({\bf r_\textrm{t}}) \psi_t^\prime({\bf r_\textrm{t}})d{\bf r_\textrm{s}} d{\bf r_\textrm{t}} .
\eeqa
This equation is equivalent to Eq.~\ref{J}, but we have explicitly included the interdot vector ${\bf R}$ in the
Coulomb operator; we assume in this case that the two variables ${\bf r_\textrm{s}}$ and ${\bf r_\textrm{t}}$ are defined from the centres of
dot I and dot II respectively. We may evaluate $V_{\rm F}$ in exactly the same way as we evaluated the intra-dot couplings,
so long as the new positions of the wavefunctions are included in the calculation. An example of the
integrand appearing in Eq.~\ref{Forster} is shown in Fig.~\ref{fig:Forstplot}. It is interesting to compare this figure to
Figs.~\ref{Jpeak} and~\ref{Kpeak}; in the plot of Fig.~\ref{fig:Forstplot} there is an extra modulation due to the extra
factor associated with the
interdot separation, and this added modulation means the integral takes longer to evaluate numerically. The results are displayed
in Figs.~\ref{fig:forster} and~\ref{fig:newforster}, where the F\"orster strength is displayed
as a function of dot separation, shape and confining potential.
The data are displayed on a log scale, and it can be seen that, for the cubic shape,
they closely follow a $1/R^3$ law for all the separations considered. This form is expected for a
dipole-dipole interaction, and we shall now discuss how
a power series expansion and subsequent approximation leads to this type of interaction in this case.
In so doing, we shall also
explain why the interaction is modified as the size and shape of the dots are changed.

\begin{figure}
\vspace{3cm}
\centerline {\hspace{-2cm}\includegraphics[bb=0 0 176 170,scale=0.9]{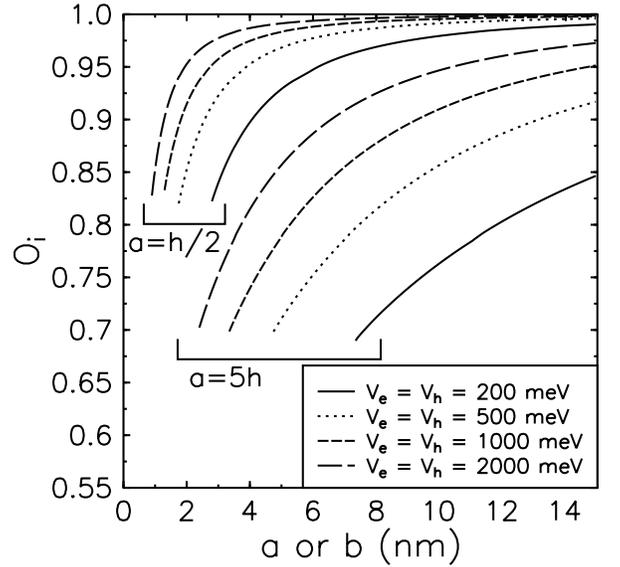}}\vspace{0cm}
\caption{The overlap integral, $O_{\rm i}$ (Eq.~\ref{overlap}), as a function of dot size and confinement potential.
This graph can be used in conjunction with Fig.~\ref{fig:forster} to obtain values of the F\"orster strength for a range
of dot sizes and confinement potentials.}
\label{fig:overlap}
\end{figure}

By making the assumption that ${\bf R}$ is much larger than ${\bf r_\textrm{s}}$ and ${\bf r_\textrm{t}}$, we may Taylor
expand the Coulomb operator. This procedure yields, to lowest non-zero order:
\begin{equation}
V_{\rm F} = \frac{C}{\epsilon_r R^3} \left(\langle {\bf r}_{\rm I} \rangle \cdot \langle {\bf r}{\rm_{II}} \rangle
-\frac{3}{R^2} (\langle {\bf r}_{\rm I} \rangle \cdot {\bf R})(\langle {\bf r}_{\rm II} \rangle \cdot {\bf R})\right)\!,
\label{dipoledipole}
\end{equation}
where it has been assumed that the dielectric constant is independent of ${\bf R+r_\textrm{s}-r_\textrm{t}}$, and as throughout the paper
is assumed to take the constant value of $\epsilon_r = 10$.
The matrix element of the position operator between an electron and a hole state on dot I or II is
\beq
\langle {\bf r}_{\rm I/II} \rangle = \int \psi_{\rm I/II}^\prime({\bf r}) \, {\bf r} \, \psi_{\rm I/II}({\bf r}) d{\bf r}.
\label{eq:posop}
\eeq
Equation~\ref{dipoledipole} is therefore
equivalent to the interaction of two point dipoles, one situated on each dot.
We can proceed further by
again employing the envelope function
approximation for electrons and holes (Eq.~\ref{envelope})
and by rewriting Eq.~\ref{eq:posop} as
\beqa
\langle {\bf r}_{\rm I/II} \rangle &=& \sum_{\{\bf T_i\}} V_{cell} \int_{cell} \phi_{\rm I/II}^\prime({\bf r-T_i})\,U^\prime({\bf r})\times \nonumber\\
&&({\bf r-T_\textrm{i}})\,\phi_{\rm I/II}({\bf r-T_\textrm{i}})\,U({\bf r})\, d{\bf r},
\eeqa
where ${\bf T_\textrm{i}}$ represents the set of lattice vectors. We have made use of the periodicity of the Bloch part of the
wavefunctions and assumed that this part of the wavefunction is the same for both dots.
By making the assumption that the envelope
function is slowly varying on the lengthscale of the atomic lattice and by using the orthogonality of the
electron and hole Bloch functions we find that:
\begin{equation}
V_{\rm F} = \frac{C}{\epsilon_r R^3} O_{\rm I} O_{\rm II}
\left(|\langle {\bf r}_{\rm a} \rangle|^2) -\frac{3}{R^2} (\langle {\bf r}_{\rm a} \rangle \cdot {\bf R})^2\right),
\label{envdipole}
\end{equation}
where the term $\langle {\bf r}_{\rm a} \rangle$ represents the atomic position operator expectation value
\beq
\langle {\bf r}_{\rm a} \rangle = \int_{cell} U_e({\bf r})\, {\bf r} \, U_h({\bf r}) \, d{\bf r},
\eeq
which is the same for both dots and
\begin{equation}
O_{\rm i} = \int_{space} \phi{\rm _e^i} ({\bf r})\, \phi{\rm _h^i} ({\bf r}) \, d{\bf r},
\label{overlap}
\end{equation}
is the overlap of electron and hole envelope functions on the appropriate dot $i$.
Eq.~\ref{envdipole} shows how the
effects of the quantum dot size and shape (which determine the overlap integrals) may be separated from the effects
of the material composition of the dot (which determine the atomic dipole operator).

It is now possible to obtain
the strength of the interaction by assuming the specific forms for the envelope and atomic functions which we discussed
earlier, in Section~\ref{model}. For the Kronig-Penney model, with a well width of $2x$,
the atomic position operator expectation value is given by
\begin{equation}
\langle {\bf r}_{\rm a} \rangle = 32x/9\pi^2.
\label{Kronig}
\end{equation}
The overlap integrals are easily calculated for the
envelope functions described earlier, and are displayed as a function of dot size and confinement potential in
Fig.~\ref{fig:overlap}. We show the overlap integral for the usual two dot shapes: cubic ($a=h/2$) and flat cuboidal
($a=5h$), where the latter is more typical of dots grown by the Stranski-Krastanow method.
The overlap is enhanced when there is a
larger confinement potential and for larger dots, since in these cases the shape of the wavefunction is less
sensitive to the effective mass difference of the electrons and holes.
We may use Fig.~\ref{fig:overlap}, together with the atomic dipole value, to calculate the
F\"orster strength for a range of dot sizes and confining potentials.
Owing to its dependence on the atomic dipole operator ($V_{\rm F}\propto x^2$),
we plot $V_{\rm F}/x^2$ as a function of $R$ in Fig.~\ref{fig:forster}.
Two example curves, for two equally sized cubic dots are shown for
equal electron and hole potentials of 500~meV and 2000~meV
in Fig.~\ref{fig:forster}, together with the earlier full calculation in which
the dipole-dipole approximation was not made. The full calculation was carried out
for (i) cubic dots with $a=h/2=2$~nm, and (ii) cuboidal dots with $a=5h=2$~nm. For both shapes it is
clear that the influence of dot shape and size is much more important in determining the size of the
interaction than the influence of the size of the confinement potentials.
Furthermore, the dipole-dipole approximation is very good in the
case of a cubic dot, even at interdot separations which are relatively small when compared to the dot sizes.
For the cuboidal dot, the dipole-dipole approximation fails at smaller separations.
We investigated this effect more thoroughly by repeating the calculations for cuboids of different
aspect ratio (see Fig.~\ref{fig:newforster}).
The next order term (dipole-quadrupole) is zero in all cases when the dots have equal size;
presumably the reason for the accuracy of
the dipole-dipole approximation in the case of cubic dots is that the dipole-dipole terms
dominate the higher order terms even at smaller dot separations.

The simple Kronig-Penney model shows how the size of the
F\"orster transfer depends upon the physical size of the atomic part of the wavefunction.
However, $\langle{\bf r}^{\rm a}\rangle$ is a widely measured quantity since it determines the strength of dipole allowed
transitions in optical spectra. In CdSe QDs it can be in the range of 0.9 to 5.2~$e$\AA,~\cite{crooker02}
in atomic systems it can also be several $e$\AA~\cite{NISTsite} and in BSs and MSs has recently been observed to be
about 1.7~$e$\AA.~\cite{hettich02}

As an illustration of the use of these curves, let us assume
that we have a dot system in which, as before, $R=5$~nm, $a=10$~nm, $b=8$~nm, and $h_1=h_2=2$~nm. Furthermore, let us
take the measured dipole value for CdSe dots of 0.9 to 5.2~$e$\AA.~\cite{crooker02}
In this case, the F\"orster strength is
between $0.013$ and $0.45$~meV, which if $\Delta_0 = 0$ would correspond to an {\it on resonance} energy transfer time of
between $318$  and $9.2$~ps in dots with $V_h=V_e=500$~meV.
This is short enough to be useful for quantum computing purposes: decoherence times
as long as a few ns~\cite{nsdeco} have been observed in QDs. In MSs or BSs, the interacting units
can be as close together as 1~nm; using this and taking a typical molecular or biomolecular dipole value of about
1.7~$e$\AA,~\cite{hettich02, crooker02} we obtain an interaction strength of 8.3~meV (or a transfer time of $\sim$497~fs).
Furthermore,  $V_{\rm F}$ must certainly be controlled if the alternative scheme  using $V_{\rm XX}$ is to be implemented
(and therefore cannot be neglected as in Ref.~\onlinecite{biolatti02}).
We note that $V_{\rm F}$ is not particularly sensitive to differences in dot size, though
the differences in the diagonal (self energy) parts of the Hamiltonian which are caused by having dots of unequal size
are very significant. We shall discuss this further in Section~\ref{conclusion}.

\begin{figure}
\vspace{4cm}
\centerline {\hspace{-2.5cm}\includegraphics[bb=0 0 171 307,scale=0.9]{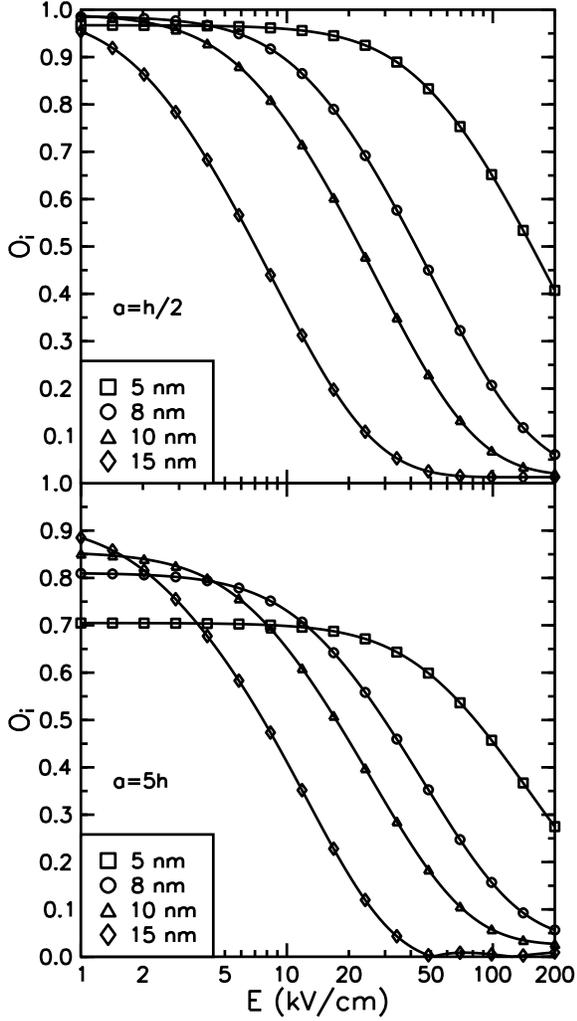}}\vspace{0cm}
\caption{The overlap integral, $O_{\rm i}$ (Eq.~\ref{overlap}), as a function of electric field strength $E$ for a
range of dot sizes.  The upper figure shows the dependence when the dots take a
cubic shape ($a=h_1/2$); the lower figure shows the dependence for a cuboidal shape ($a=5h_1$).
Note that the overlap integral, and so also the F\"orster interaction, is suppressed at large field as
the electron and hole are forced apart.}
\label{fig:OLfielda}
\end{figure}

\begin{figure}
\vspace{4cm}
\centerline {\hspace{-2.5cm}\includegraphics[bb=0 0 171 307,scale=0.9]{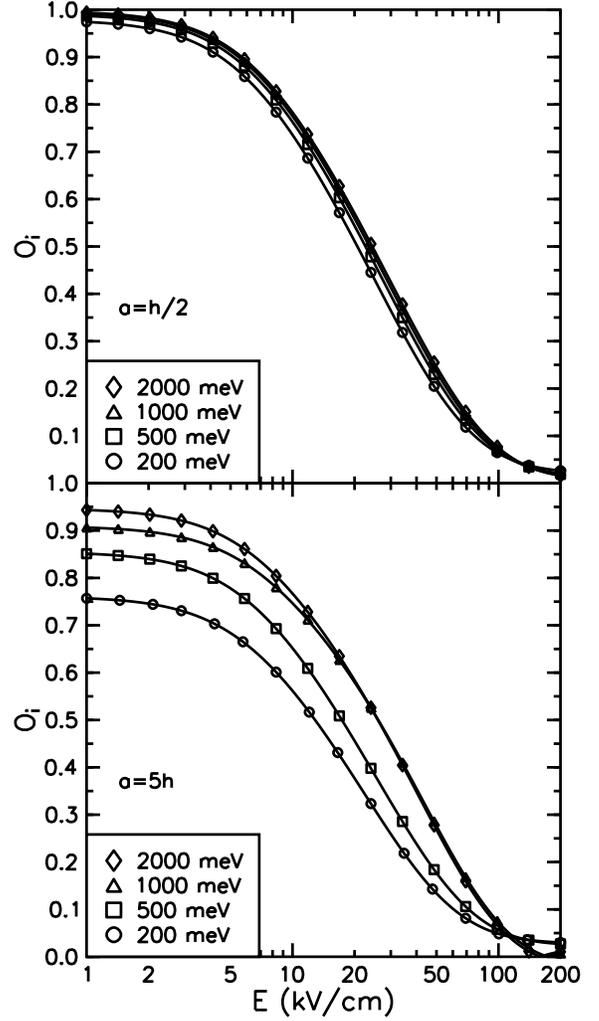}}\vspace{0cm}
\caption{The overlap integral, $O_{\rm i}$ (Eq.~\ref{overlap}), as a function of electric field strength $E$ for a
range of confinement potentials, for a dot size of $a=10$~nm. The upper figure shows the dependence when the dots take a
cubic shape ($a=h_1/2$); the lower figure shows the dependence for a cuboidal shape ($a=5h_1$).}
\label{fig:OLfieldpot}
\end{figure}

In the Section~\ref{VXX} we shall discuss how the biexciton binding energy term depends upon applied electric field, and
how such a field may be crucial to the operation of a potential quantum logic device. We now discuss
how the F\"orster term would vary when such a field is applied. Since an electric
field would move the electron and hole away from one another, the overlap integrals, Eq.~\ref{overlap}, would be reduced
by such a field. We show this specifically by simulating the effect of applying a field on the overlap integral $O_i$ in
Figs.~\ref{fig:OLfielda} and \ref{fig:OLfieldpot}. In Fig.~\ref{fig:OLfielda}, we see that $O_i$ is significantly suppressed
for fields of a few 10's of kV/cm, and that the suppression is easier to achieve in dots which have a larger dimension
in the direction of the applied field. The reason for this is that in larger dots the electron and hole are more easily
separated since there is more distance between the two dot-barrier interfaces. In zero field, dots with a
{\it smaller} dimension have smaller overlap integrals, for the reasons associated with the balance of kinetic and potential
energy discussed earlier. Hence, we see that the curves for different dot sizes cross each other in an applied field.
In Fig.~\ref{fig:OLfieldpot} we plot the dependence of the $O_i$ on the depth of the confinement potentials, for $a=10$~nm
for both the cubic and cuboidal geometries. We see here that
the effect of varying confinement potential is much smaller than varying dot size. The small difference that is evident,
that of a slightly easier suppression for deeper confinement potentials, is presumably due to the fact that the
wavefunctions in a shallower wells tend to be more spread out (since the potential energy cost in doing so is smaller),
and so
the overlap between an electron and hole at opposite sides of the dot is slightly enhanced. Again the curves cross one
another
since, as we discussed earlier, in zero field the delocalization of the states for shallower potentials means that the
shape of each particle's wavefunction depends more strongly on its effective mass.

The fact that the F\"orster coupling may be suppressed by an external field could be very useful: if an entangled state
is produced by using this coupling, it may be maintained by switching off $V_{\rm F}$. If this could be done in a
sufficiently
short time (i.e. on the timescale of the evolution of the quantum device under the F\"orster coupling Hamiltonian, but much
less than typical decoherence times), it may be possible to fabricate a two qubit gate using this effect. It may
also be possible to achieve this switching in an alternative way, by leaving the F\"orster coupling at a constant
value but by tuning the single exciton level spacing $\Delta_0$ through the electric field which induces the Stark
shift;~\cite{harrison00} we shall return to this in Section~\ref{conclusion}.
It is also interesting that a negligible F\"orster coupling is essential for the energy selective dot device
discussed in Section~\ref{heisenberg}: such a negligible coupling may be achieved through using an external field.
There are disadvantages doing this in this case, however. A smaller electron-hole overlap will also reduce the coupling to
the light field itself, which we need to be strong enough to be able to perform conditional gates in a time short enough
when compared with typical decoherence times. Hence, as so
often in quantum computing implementation, a compromise must be struck between these two requirements.

\subsection{Diagonal coupling}
\label{VXX}
We now calculate the direct Coulomb interaction
between two excitons, where one exciton is located on each dot. This interaction leads to the
energy selectivity of the gate and is the responsible for the $V_{\rm XX}$ term of Eq.~\ref{eq:Hdot}.

Consider the following double substitutional Slater determinant, which we write using the hole prescription described in
Section~\ref{coulomb}. It
represents a combination of two ground conduction electron states and two ground hole states which correspond to one exciton
on each dot:
\beq
\Psi_{\rm XX} = \cal{A}\Big[\psi_{\rm e}^{\rm I}({\bf r_\textrm{1}})\psi_{\rm h}^{\rm I}({\bf r_\textrm{2}})\psi_{\rm e}^{\rm II}({\bf r_\textrm{3} - R})
\psi_{\rm h}^{\rm II}({\bf r_\textrm{4} -  R})\Big],
\label{XXwf}
\eeq
where $\cal{A}$ indicates that the wavefunction has overall antisymmetry, this being achieved by
adding terms with labels swapped around in a Slater determinant form. ${\bf R}$ is the vector connecting the two dot
centres, ${\bf r_\textrm{1}}$ and ${\bf r_\textrm{3}}$ represent the position vectors of electrons relative to the centres of
dot I and dot II respectively and ${\bf r_\textrm{2}}$ and ${\bf r_\textrm{4}}$ are the equivalent vectors for holes.
The associated Coulomb operator $\hat{V}_{\rm XX}$ is given by
\beqa
\hat{V}_{\rm XX} &=& \frac{C}{\epsilon_r}\left[\frac{1}{|{\bf R}+{\bf r_\textrm{1}} - {\bf r_\textrm{3}}|}-\frac{1}{|{\bf R}+{\bf r_\textrm{1}} - {\bf r_\textrm{4}}|} \right.  \nonumber \\
                  && \left. -\frac{1}{|{\bf R}+{\bf r_\textrm{2}} - {\bf r_\textrm{3}}|} +\frac{1}{|{\bf R}+{\bf r_\textrm{2}} - {\bf r_\textrm{4}}|}\right].
\eeqa
Expanding this expression in a Taylor series about ${\bf R}$ gives, to lowest non-zero order:
\beq
\hat{V}_{\rm XX} = \frac{C}{\epsilon_rR^3}\left\{{\bf p_{\rm I}}\cdot{\bf p_{\rm II}} -
\frac{3}{R^2}({\bf p_{\rm I}}\cdot{\bf R})({\bf p_{\rm II}}\cdot{\bf R})\right\}\! ,
\label{XXdipole}
\eeq
where ${\bf p_{\rm I}} = e({\bf r_\textrm{1}} - {\bf r_\textrm{2}})$ is the overall dipole
moment on dot I
and ${\bf p_{\rm II}} = e({\bf r_\textrm{3}} - {\bf r_\textrm{4}})$ is the overall dipole moment on dot II.
To evaluate
the matrix element $\langle \Psi_{\rm XX}|\hat{V}_{\rm XX}|\Psi_{\rm XX}\rangle$,
${\bf p}_{\rm I}$ and ${\bf p}_{\rm II}$ in Eq.~\ref{XXdipole} are replaced by their expectation values for the wavefunction,
Eq.~\ref{XXwf}. This procedure gives rise to
a direct term and exchange terms. The exchange terms arise from the
parts of the wavefunction (Eq.~\ref{XXwf})
which do not appear explicitly within the bracket but which have their labels swapped around and they
are zero in the absence
of wavefunction overlap between dots. The direct term is obtained through the use of the envelope function
approximation, Eq.~\ref{envelope},
which leads to the following equation for the expectation value
$\langle {\bf r_\textrm{1}} \rangle$
\beq
\langle {\bf r_1} \rangle = \int_{space} \phi_{\rm e}^{\rm I\ast} ({\bf r_\textrm{1}})\, {\bf r_\textrm{1}}\, \phi_{\rm e}^{\rm I} ({\bf r_\textrm{1}})\, d{\bf r_\textrm{1}},
\label{xop}
\eeq
where the orthogonality of the Bloch functions for different bands and the slow variation approximation for the
envelope functions have again been used.
Similar expressions hold for the other position expectation values.

For a cubic dot, where
the electron and hole wavefunctions have a definite parity about the dot centre, Eq.~\ref{xop} implies that
the exciton-exciton coupling is zero. However, this is
not the case when this symmetry is broken.
For instance, for pyramidal shapes
the electron may localize in one region of the dot and the hole in another region.~\cite{pryor}
Alternatively, an electric field
would induce a polarization on the dot; this field may be externally applied or arise from intrinsic
piezoelectric effects.~\cite{biolatti02,derinaldis02}

\begin{figure}
\vspace{4cm}
\centerline {\hspace{-3cm}\includegraphics[bb=0 0 123 220,scale=1.2]{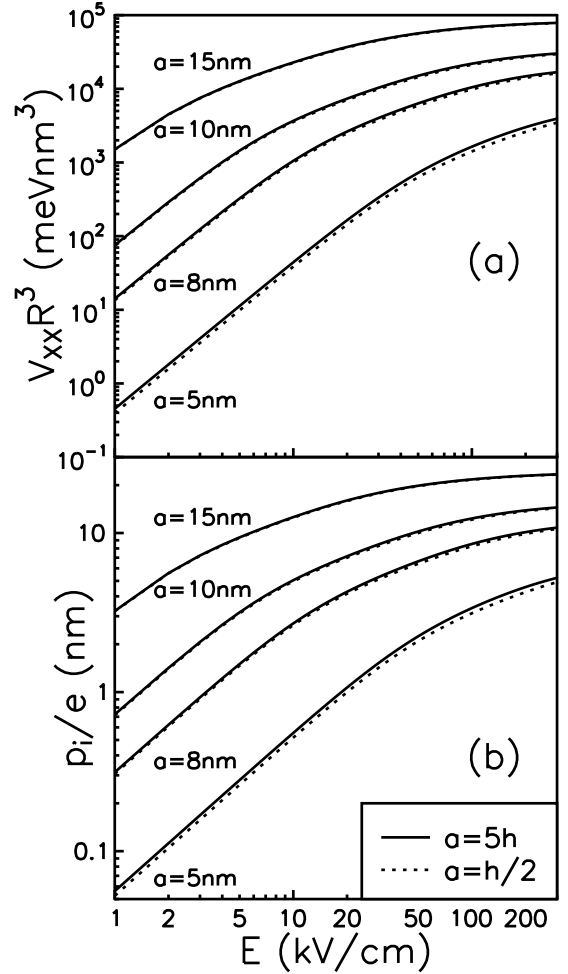}}\vspace{0cm}
\caption{(a) Exciton-exciton binding energy and (b) induced dipole moment
as a function of the dot size, shape and applied electric field. We have assumed
that $V_e = V_h = 500$~meV.}
\label{fifig}
\end{figure}

We have simulated the effect of applying an electric field in our cuboidal model by including a
linear potential in the single particle Schr\"odinger Eqs.~\ref{eq:spschr},
and the results are displayed in Fig.~\ref{fifig}. In the lower part of the figure we display the size of the exciton
dipole moment
$p_{\rm i}$ on one of the dots as a function of the dot size and of the applied electric field strength $E$.
We do this for our usual two geometries: cubic ($a=h_1/2$) and flat cuboid ($a=5h_1$) (see Fig.~\ref{dotspic}).
In both cases the field is applied along one of the axes of the square base.
The interaction strength $V_{\rm XX}$ is then obtained by using the size of
$p_{\rm i}$ for each dot and substituting into Eq.~\ref{XXdipole}.
Thus, the upper part of Fig.~\ref{fifig} shows the strength $V_{\rm XX}$, normalized by  $R^{3}$, for two dots of
equal size and calculated for both of the dot geometries just described. (The interaction between two unequally sized dots
can similarly be obtained by the use of Fig.~\ref{fifig}(b) and Eq.~\ref{XXdipole}.)
At very small applied field, the induced dipole is linearly proportional to the field, and hence the interaction strength
takes a quadratic dependence on field. At larger applied fields,
the induced dipole begins to saturate as the electron and hole approach the edges of confining potential of
the quantum dots; this limits the useful interaction strength which may be obtained from a given pair of dots.
It is interesting that the interaction strength is much more dependent on $a$ than it is on $h_1$; this is because
it is in the basal plane that the field is applied, and so it is in this direction that the dipole moment is induced.
The relative insensitivity of $V_{\rm XX}$ to $h_1$ turns out to be very useful: it means that SK dots, which can be stacked
closely on top of one another but which have a relatively large base size, can be made to interact very strongly.
As an illustration, consider the following typical SK dot parameters: $R=5$~nm, $a=8$~nm,  $b=10$~nm, and $h_1=h_2=2$~nm
(we have assumed the usual experimental situation in which the upper dot of the stack has a slightly larger size).
In an $x$ directed field of 100~kV/cm, these parameters give $V_{\rm XX} \approx 120$ meV, which
would result in a  lower time limit for the gate operation
of around 10~fs. This is relatively short; decoherence times on the order of nanoseconds have been observed for
uncoupled dots.~\cite{nsdeco}
Finally, we note that we have only calculated $V_{\rm XX}$ to first order; in some cases higher order terms may be
important.

\section{Further Discussion}
\label{conclusion}

The model outlined above can also be used to calculate $\Delta_0$,
the difference in exciton creation energy
for two different sized dots in the absence of interactions. This is done by simply calculating the
single particle electron and hole energy and taking into account the Coulomb binding energy between them.
We also assume that the electron and hole are in the spin singlet state; spin is not important for our present
proposal and it is always possible to choose the spin singlet state by using light with the appropriate polarization.
Hence we can effectively ignore the triplet states in considering the scheme described in Section~\ref{heisenberg}.

\begin{figure}
\vspace{4cm}
\centerline {\hspace{-2.5cm}\includegraphics[bb=0 0 173 307,scale=0.9]{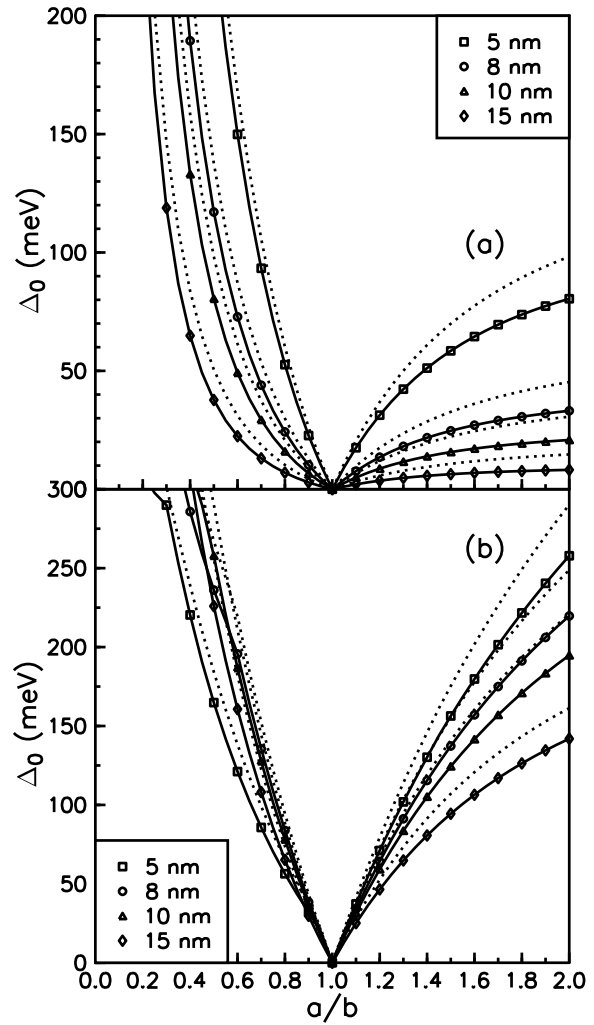}}\vspace{0cm}
\caption{Energy splitting $\Delta_0\equiv \w_1-\w_2$ of the singlet qubit exciton states $\ket{\Psi_{01}}$,
and $\ket{\Psi_{10}}$ for (a) cubic dots ($a=h_1/2, b=h_2/2$) and (b) cuboidal dots ($a=5h_1, b=5h_2$) in the absence of
the F\"orster interaction as a function of the dot size ratio $a/b$.
The splitting is independent of interdot distance. The solid lines represent the splitting in the presence of the
Coulomb and exchange splitting terms and each adjacent dotted line represents the splitting without Coulomb interactions
for each dot size.}
\label{deltazero}
\end{figure}

\begin{figure}
\vspace{4cm}
\centerline {\hspace{-2.5cm}\includegraphics[bb=0 0 173 307,scale=0.9]{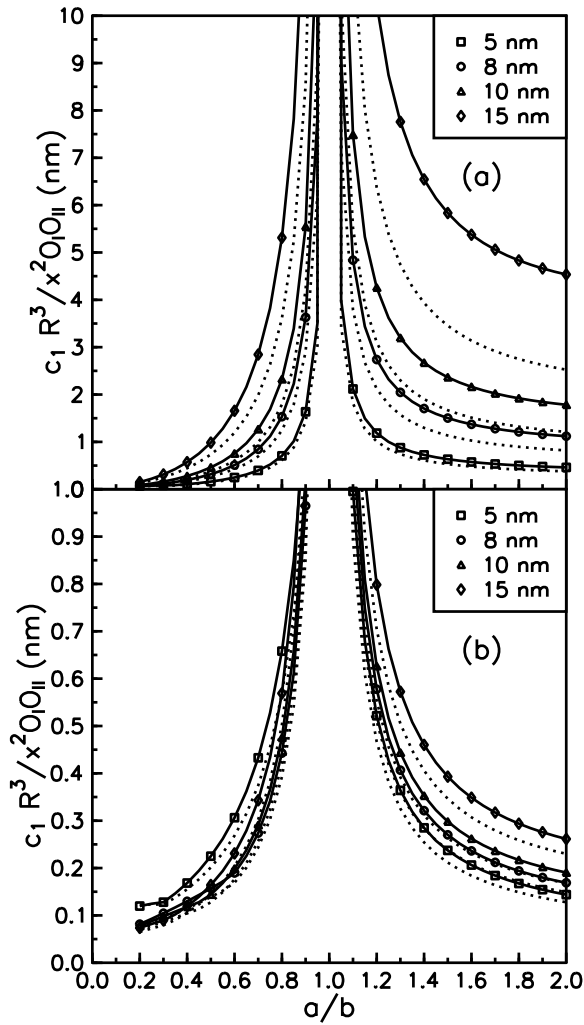}}\vspace{0cm}
\caption{The size of the component of the wavefunction $c_1$ for (a) cubic dots ($a=h_1/2, b=h_2/2$) and
(b) cuboidal dots ($a=5h_1, b=5h_2$)
as a function of the dot size ratio $a/b$.
$c_1$ has been scaled by its dependence on the interdot distance, $R$, typical atomic spacing, $x$, and
overlap integrals, $O_i$.}
\label{cdep}
\end{figure}

The absolute value of $\Delta_0$ is displayed as a function of the ratio of the dot side lengths for each of the usual
geometries in Fig.~\ref{deltazero}.
We have displayed $\Delta_0$, both in the absence of the (intra-dot) Coulomb terms, and when these terms are included
to first order by using the calculations of Section~\ref{intra}. $\Delta_0$ is zero when the dots are of equal size, and
then
increases as the difference in size becomes greater. The Coulomb terms serve to reduce the size of $\Delta_0$ because
they are larger for the smaller dot of the pair (which also has the larger single particle energies).

We now use the analysis of Section~\ref{heisenberg} and the calculations of subsequent sections to obtain
the size of the $c_1$ component of the $\ket{\Psi_{10}}$ and $\ket{\Psi_{01}}$ states.
If we assume that the F\"orster strength is small in comparison with $\Delta_0$,
then we have that
$c_1 \approx V_{\rm F}/\Delta_0$ (see inset of Fig.~\ref{QC}(c)) and by substituting Eq.~\ref{Kronig} and Eq.~\ref{envdipole}, we obtain
\begin{equation}
\frac{R^3c_1}{x^2O_{\rm I} O_{\rm II}} \approx \frac{37.1}{\Delta_0} \, ,
\end{equation}
where $\Delta_0$ is measured in meV, and $x$ and $R$ are in nm. This quantity is displayed as a function of dot size
ratio for the usual geometries in Fig.~\ref{cdep}.
It can be seen there that a range of $c_1$ values can be obtained by choosing dots with appropriate values of
$x$, $R$ and $a/b$. For example, cubic dots with large $x$ ($>1$~nm say), small $R$ ($<3$~nm say) and $a/b \sim 1$
give a larger $c_1$,
and it is then more appropriate to use the F\"orster interaction itself to create entangled states.
On the other hand, dots with smaller $x$,
larger $R$ or a large mismatch in dot size would be more suited to the scheme which uses the
$V_{\rm XX}$ for QC and entanglement generation. A scheme similar  to the latter one  was discussed by Biolatti {\it et al.}
in Ref.~\onlinecite{biolatti02}, though the off-diagonal coupling was not considered there at all; we now see how important
it is to consider the effect of this interaction.
The {\emph {fidelity}}~\cite{poyatos97} of a typical $V_{\rm XX}$ entangling gate operation (e.g. $\ket{11}
\mapsto \ket{10}$) is equal to $1-c_1^2$---and so one must be
careful when using the biexciton scheme
to use the available parameter space and make sure that the F\"orster transfer is suppressed to the desired accuracy.
There are other sources of decoherence in this case (e.g. the interaction with optical and acoustic
phonons~\cite{jhthesis,nsdeco,reina02}) which will reduce the value of the fidelity to below $1-c_1^2$.
To minimize the effects of such decoherence channels, it is important to maximize $V_{\rm XX}$,
since this leads to an improved transition discrimination and so to a faster gating time.
This can be done by applying an electric field and choosing an appropriate dot shape, size
and separation (as described earlier).
It is then necessary to minimize the basis state mixing for the chosen parameters by selecting a
suitable dot size ratio and material composition.
It was seen earlier that the value of $V_{\rm XX}$ could be as high as several tens of meV. If we assume a conservative
figure of
10~meV, we find that the uncertainty principle implies that a CNOT gate could be performed in a time of around 100~fs.
This is
relatively short; decoherence times on the order of nanoseconds have been observed recently for
uncoupled dots.~\cite{nsdeco} Hence we conclude that this scheme looks rather promising as a solid state implementation of
quantum computation.

It is also possible to use the Stark effect~\cite{harrison00} to tune two non-resonant levels into resonance,
thereby allowing for the kind of switching of the F\"orster interaction which is required if it is to be used
for quantum logic. This can be done so long as
the two dots are made such that they have different polarizibilities (which could be achieved by using either different
sized dots or dots made of different materials). So long as this difference is such that the levels are brought closer
together by applying a field, and that shifts as large as $\Delta_0$ can be achieved, switchable resonant transfer is
possible. It might be difficult to achieve the switching in a time which is short enough for a quantum gate to be
performed; however, in this case the optical (AC) Stark shift could be employed by using ultrafast lasers.~\cite{cohen2}

Single shot qubit state measurement in QDs could be performed by using resonant fluorescent shelving
techniques.~\cite{blatt99}
The QD state measurement can also be achieved by means of projecting onto the computational basis and measuring the
final register state by exploiting ultrafast near-field
optical spectroscopy and microscopy:~\cite{gammon01,schelpe02} these allow one  to address, to excite and to probe
the QD excitonic states with spectral and spatial selectivity.  In addition,
the qubit  register density matrix can be reconstructed by measuring
the QD photon correlations via standard quantum state tomography techniques.~\cite{bill02s}
In particular, we believe that the activity of the (F\"orster) resonant energy transfer processes discussed in this paper
can be accomplished in our coupled dot molecule  by measuring the intensity correlation function
(usually denoted $g^{(2)}$) in a
Hanbury-Brown/Twiss type experiment.~\cite{rege01,santo01,berglund01}
Such an experiment can reveal signatures of purely non-classical
photon correlations arising from the QD molecule emission
(i.e., photon antibunching or bunching behaviour).
This idea has already been experimentally explored for the case
of pairs of dye molecules by Berglund {\emph {et al.}}.~\cite{berglund01}
Scalability of the scheme given here could also be possible by adopting a globally addressed qubit
strategy~\cite{benjamin00} on a stack of self-organized QDs.~\cite{xie95}

We finalize this section with a discussion on how the $V_\textrm{F}$ coupling can be usefully manipulated in biomolecular nanostructures.
Light-harvesting antenna complexes~\cite{hu02} or arrays of strongly interacting individual molecules~\cite{hettich02} could
provide an appropriate system in which the F\"orster interaction could be
used for QIP tasks. They are generally very uniform structures,
and we may compare them to QDs by setting $a/b \sim 1$, or
$V_{\rm F}/\Delta_0\gg 1$. Then the one-exciton eigenstates of a two qubit system with a F\"orster coupling naturally
allows the generation of the states $\raiz (\ket{01}\pm\ket{10})$, which, apart from their applications to quantum
protocols, can be particularly useful in the fight against decoherence. Spectroscopic, line-narrowing techniques
(e.g., hole burning and site-selective fluorescence), infrared and Raman experimental studies reveal that the main
decoherence mechanisms in  the antenna complexes arise from energetic disorder, electron-phonon coupling, and temperature
effects.~\cite{hu02}  In this scenario, the excitations couple to an environment that typically possesses a much larger
coherence length than the biomolecular units (BChl's) spacing. For example, the BChl's in the antenna complex LH2, which we regard as a potential system for quantum logic, are spaced by as little as 1~nm, and hence so-called collective decoherence is
expected to apply. In this case, provided that the logical qubit encoding $\downket_i\equiv\ket{01}_{jk}$,
$\upket_i\equiv\ket{10}_{jk}$ that uses two physical (exciton) qubits can be realized in the BChl's system,
arbitrary superpositions of  logical
qubits such as $(\alpha_i\downket_i+\beta_i\upket_i)^{\otimes N}$, $i=1,\ldots, N$, $\alpha_i$, $\beta_i\in\mathbb{C}$,
are immune to dephasing noise (described by a $\sigma_z$ operator~\cite{reina02}), and single qubit manipulations can be
carried out on the timescale of the
F\"orster coupling (which as we have seen can be as short as 497~fs). Two-qubit logic gates can also be implemented
within  a decoherence-free subspace by using the above encoding,~\cite{lidar98} thus completing a universal set of gates.
Initialization of the system requires the pairing of the physical qubits to the logical `ground' state
$\ket{\downarrow}_i^{\otimes N}$, and  readout is to be accomplished by  identifying on which of the two structures the
exciton resides. Furthermore, rings of BChl's appear side by side in naturally occuring antenna complexes
and also display energy selectivity---smaller rings tend to
have higher energy transitions.~\cite{hu02} Thus, following a scheme as above,  it may be possible to scale up
such biological units in a natural way and construct a robust energy selective scheme for quantum
computation.~\cite{reina03} We also note that arrays of strongly interacting individual molecules which are coupled
via a near field dipole-dipole interaction are well suited for our quantum computing and entangling schemes, especially
due to the existence of $V_{\textrm{XX}}$-type of energy shifts,~\cite{hettich02} which have been analyzed in this paper.
A full discussion of these ideas will be presented elsewhere.~\cite{reina03}

\section{Summary}

We have shown that it is possible to use the two different electrostatic coupling terms (F\"orster transfer
and biexciton binding energy) between excitons in quantum dots to construct two
qubit gates which, in addition to an appropriate control over single qubits, are enough for universal quantum computation.
We have also discussed how to generate tailored exciton entangled states by using these gates.
We have
furthermore modelled a pair of quantum dots in the simplest envelope function approximation, and have
mapped out the areas of parameters space where one of the two interactions dominate. We have discussed in detail how to
perform entangling operations when one of the two interactions is dominant;
the case when the two have similar magnitude leads to a rich spectrum of entangled states whose degree of
entanglement can be quantified {\it a posteriori} in, for example, photon correlation experiments.
We have concentrated here on two geometries, namely that of a cube-shaped dot and that of a cuboid-shaped dot,
and our calculations have been partially analytical and partially numerical. In future, we are hoping to obtain simpler
analytical results for many of the quantities we have calculated, by using different dot geometries. We shall also
extend our calculations to include decoherence effects, and in future work we shall be particularly interested on how we
might use the F\"orster interaction to create and use decoherence free subspaces.

\section{Acknowledgements}
We would like to thank R.~A.~Taylor, S.~C.~Benjamin, S.~D.~Barrett, R.~G.~Beausoleil, T.~P.~Spiller and
W.~J.~Munro for useful and stimulating discussions.
The project is supported by the Foresight Link Award {\it Nanoelectronics at the Quantum Edge}.
BWL is grateful to St Anne's College for financial support.

\appendix{}
\section{Single Particle Solutions}
As we discussed in the text, the solution of the three dimensional finite well square box potential is obtained by using
the solutions of the one dimensional finite square well and expanding the Schr\"odinger equation in these basis states.
In order to do this, both bound and unbound basis states had to be taken into consideration.

\subsection{Bound States}
\begin{figure}
\vspace{5cm}
\centerline {\hspace{-2cm}\includegraphics[bb=0 0 391 401,scale=0.35]{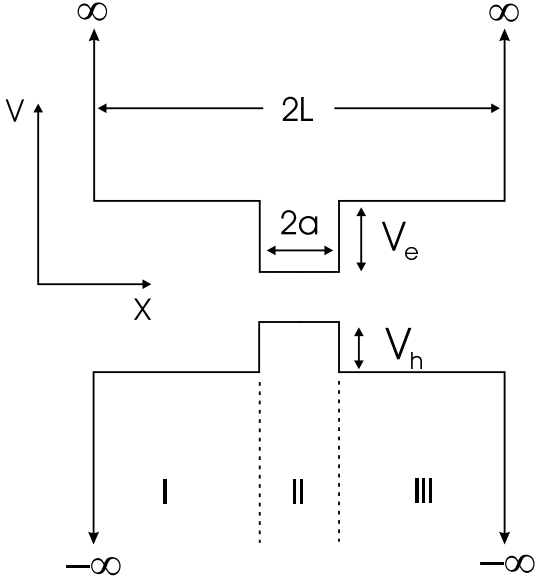}}\vspace{-2.5cm}
\caption{Schematic diagram of the potential used to generate the unbound basis states used in the calculations.}
\label{unboundfig}
\end{figure}

The problem of finding the solutions of a finite square well is covered in most undergraduate text books
(see e.g. Ref.~\onlinecite{cohen}), and so we shall not go into too much detail here. If we assume that the finite well is
centred around $x=0$, we have different forms of solution depending on the parity of the wavefunction. With reference
to Fig.~\ref{unboundfig}, and assuming that our wavefunctions have decayed once the infinite barriers are reached
(see Section~\ref{spstates}), we have

\begin{table}[h]
\begin{center}
\begin{tabular}{|c|c|c|} \hline
Region          & Even Solutions  & Odd Solutions \\  \hline
$x<-a/2$       & $Ae^{\alpha x}$   &  $Ce^{\alpha x}$    \\
$-a/2<x<a/2$  & $B\cos(kx)$       &  $D\cos(kx)$        \\
$x>a/2$      & $Ae^{-\alpha x}$  &  $-Ce^{-\alpha x}$  \\ \hline
\end{tabular}
\caption{Table of the forms of the bound state solutions for the finite square well potential.}
\end{center}
\label{tab:bound}
\end{table}

Where A, B, C and D are normalization constants. $\alpha = (2m_p^\ast (V_p-E))^{\frac{1}{2}}/\hbar$
and $k = (2m_p^\ast E)^{\frac{1}{2}}/\hbar$. By ensuring the continuity of the amplitudes of the wavefunctions
and the probability current at the boundaries, the following transcendental equations for the energy $E$ are obtained:

For even solutions:
\begin{equation}
\tan\left(\frac{(2m_p^\ast E)^{\frac{1}{2}}a}{2\hbar}\right) = \left(\frac{V_p-E}{E}\right)^{\frac{1}{2}}.
\end{equation}

For odd solutions:
\begin{equation}
\cot\left(\frac{(2m_p^\ast E)^{\frac{1}{2}}a}{2\hbar}\right) = -\left(\frac{V_p-E}{E}\right)^{\frac{1}{2}}.
\end{equation}
These equations were solved by using the numerical root finding algorithm provided with the NAG package. Once the
energies are obtained, $\alpha$ and $k$ follow from simple substitution. The normalization constants follow from
the usual normalization procedures once all of the other parameters are known.

\subsection{Unbound States}
The problem for unbound states is somewhat less straightforward.
We assume that the finite well (of width $2a$) is embedded within an infinite well (of width $2L$), and that we may set the
wavefunction outside the infinite well to zero (see Fig.~\ref{unboundfig}).
Then there are again three regions which have different forms of solution, which take the form:
\begin{table}[h]
\begin{center}
\begin{tabular}{|c|c|c|} \hline
Region          & Even Solutions  & Odd Solutions \\  \hline
$-L/2<x<-a/2$  & $-A\sin(k^\prime(x-L))$   &  $ C\sin(k^\prime(x-L))$   \\
$-a/2<x<a/2$  & $ B\cos(kx)$              &  $ D\cos(kx)$             \\
$a/2<x<L/2$  & $ A\sin(k^\prime(x-L))$    &  $-C\sin(k^\prime(x-L))$   \\ \hline
\end{tabular}
\caption{Table of the forms of the unbound state solutions for the finite square well potential.}
\end{center}
\label{tab:unbound}
\end{table}

The wavevector inside the dots is given by $k=(2m^\ast E)^\frac{1}{2}/\hbar$, and
$k^\prime=(2m^\ast (E-V_p))^\frac{1}{2}/\hbar$.
The solution is obtained by again invoking the continuity of the wavefunction amplitude and the probability current  at
the boundaries, and the following transcendental equations are found for the even and odd solutions respectively:
\beqa
\label{unboundeven}
\left(\frac{E-V_p}{E}\right)^\frac{1}{2} &=&
-\tan\left(\frac{a(2m^\ast E)^\frac{1}{2}}{\hbar}\right)\times \\ \nonumber
&& \tan\left(\frac{(a-L)(2m^\ast (E-V_p))^{\frac{1}{2}}}{\hbar}\right)\, ,
\eeqa

\beqa
\label{unboundodd}
\left(\frac{E-V_p}{E}\right)^\frac{1}{2} &=&
\cot\left(\frac{a(2m^\ast E)^\frac{1}{2}}{\hbar}\right)\times \\ \nonumber
&& \tan\left(\frac{(a-L)(2m^\ast (E-V_p))^\frac{1}{2}}{\hbar}\right) \, .
\eeqa

Eqs.~\ref{unboundeven} and~\ref{unboundodd} are solved for $E$, and then the wavevectors and normalization
constants follow on as before.


\end{document}